\documentclass{article}[11]
\usepackage{amsmath}
\usepackage{graphicx}
\usepackage{amsmath, amssymb,amsthm}
\usepackage{xcolor} 
\usepackage{chngcntr}

\begin{document}

\date{}

\title{Tests on High-Directivity Unconventional Biconical Type Antennas}
\maketitle

\begin{center}
  Daniele Funaro\\
  {\small Dipartimento di Scienze Chimiche e Geologiche \\
    Universit\`a  di Modena e Reggio Emilia, Via Campi 103, 41125 Modena, Italy}

  {\small daniele.funaro@unimore.it}
\end{center}

\begin{center}	
  Alessandro Chiolerio \\
  {\small Center for Bioinspired Soft Robotics \\ Istituto Italiano di Tecnologia, Via Morego~60, 16165~Genova,~Italy}

  {\small alessandro.chiolerio@iit.it}
\end{center}

\begin{abstract} 
Biconical-type antennas featuring high directivity have been designed, created, and tested in
anechoic chamber. Results in the range between 1 and 5 GHz are presented in this article. In particular,
two different configurations have been tested,  with and without dielectric lenses, both involving  rapid
prototyping tools (3D printing) for the dielectric and the antenna support. A very high directivity is
nowadays demanded by efficient and sustainable point-to-point communications or energy transfer protocols,
to avoid releasing energy in neighboring areas and preserve data transfer security. As demonstrated here, special biconical type antennas featuring a
3D printed polylactic acid (PLA) dielectric lens can achieve a good directivity, with a corresponding
emission lobe centered around $8.4$ degrees, featuring a FWHM of
$6.4$ degrees. Dielectric lens-free antennas, featuring an unconventional shape,  can also achieve a good directivity, with a 
corresponding emission lobe centered around $10.0$ degrees, featuring a FWHM of
$14.2$ degrees. The preliminary results shown here explore some of the aspects of the vast configuration space (which include
fabrication techniques, dielectric materials, conductive supports, etc.) and open the route for further
optimization studies. The aim would be to adjust the various degrees of freedom in order to achieve
what can be defined as ``infinite'' directivity.
\end{abstract}

\vspace{.8cm}
\par\smallskip

\counterwithout{figure}{section}
\counterwithout{figure}{subsection}

\section{Introduction}\label{sect:Intro}
Highly directive antennas are an important
tool in all applications where point-to-point communications need to be established. 
There are several reasons to
justify their development, as for example to guarantee secure
transmissions, to reduce the impact of electromagnetic fields
on neighbouring areas or the interference between spatially
distributed components, to increase the sustainability and
therefore damp transmission losses in particular when a high
power is radiated \cite{Shukla}, \cite{Fantuzzi}. There are currently two categories of
commercial directive radio antennas \cite{Georgiou}: on the one hand we
find the traditional ones, including helix, log-periodic array,
aperture horn, reflectors and patch antennas. On the other
hand there are smart antennas, which have become known as
“beam-forming” \cite{Pandey}, consisting of an array of elements that
can offer adaptive transmission. Optical means are also used
to convey directional channels \cite{Khalighi}. A recent use of dielectric
lenses to correct the direction of the signal emitted is reported
in \cite{ding}. Depending on the specific application, shapes, sizes
and designs can be quite different. The polar signal pattern
from a directional antenna typically consists of an elliptic
principal lobe, usually surrounded by smaller side lobes. For
a survey on radiation patterns and an application on wireless
networks we mention for instance: \cite{DaiNgLiWu}, \cite{Hurley}.

Biconical antennas have been used ever since in the transmission range between 30 and 200 MHz and their theory is
known since the 1950s \cite{Schelkunoff}. Extensions up to 1 GHz \cite{tanaka} and
Ultra Wide Band applications (UWB) \cite{black}, have been recently
implemented. Biconical antennas have been adopted in UWB
Direction of Arrival (DoA) applications \cite{remez}. Asymmetric
designs have also shown to provide beam shaping capabilities
\cite{ghalibafan}.
In space applications biconical antennas are also used (see,
e.g., \cite{shi}), particularly for the low or medium gain segments,
as reported for example in \cite{vacchione}, where the directivity achieved
by a corrugated design shows a Full Width at Half Maximum
(FWHM) of about $50^\circ$. In this last case, the finality was to
achieve a broad beam, which is exactly the opposite of what
we would like to get here.

The results reported in the present paper come from the
practical realization of the device proposed and designed in
\cite{chiolerio}, where numerical simulations were also presented. The
main peculiarity of this typology of antenna is that the signal
departing from the emitter is composed by successive annular
regions, traveling from the coaxial cable (TEM-mode) and
emerging always preserving the same topology. A theory that
predicts the existence of “solitary” waves that travel in free-space along straight patterns without dissipation,
has been developed \cite{Fun2}, \cite{Fun3}. This included the
case in which the Poynting vectors are all parallel. Such an “infinite” directivity may
be not achievable in practice. Nevertheless, starting from the
preliminary test here reported, it is certainly possible to design
more evolved prototypes that reproduce at their best such an
extreme behavior.
 
\medskip

 {\bf Acknowledgements} For their kind assistance, we are indebted with Prof E.
Bassoli (3D printing of the versions C and D) and Prof. F.
Leali (preparation of the aluminium reflector), Dipartimento di
Ingegneria Enzo Ferrari, Universit\`a degli Studi di Modena e 
Reggio Emilia; Prof F. Auricchio (3D printing of the dielectric
lenses), 3D@UniPV, Department of Civil Engineering and
Architecture, Universit\`a degli Studi di Pavia; Prof. L. Ferraris, 
Dipartimento di Energia, Politecnico di Torino. Special thanks
for the valuable support in the measurements process go to Dr.
F. Franchini, Laboratorio di Compatibilità Elettromagnetica, 
Politecnico di Torino, located in Alessandria.

\section{Biconical antennas with 3D printed dielectric}\label{sect:biconic}
%

\begin{figure}[h!]
\centering
\includegraphics[width=9.6cm,height=5.9cm]{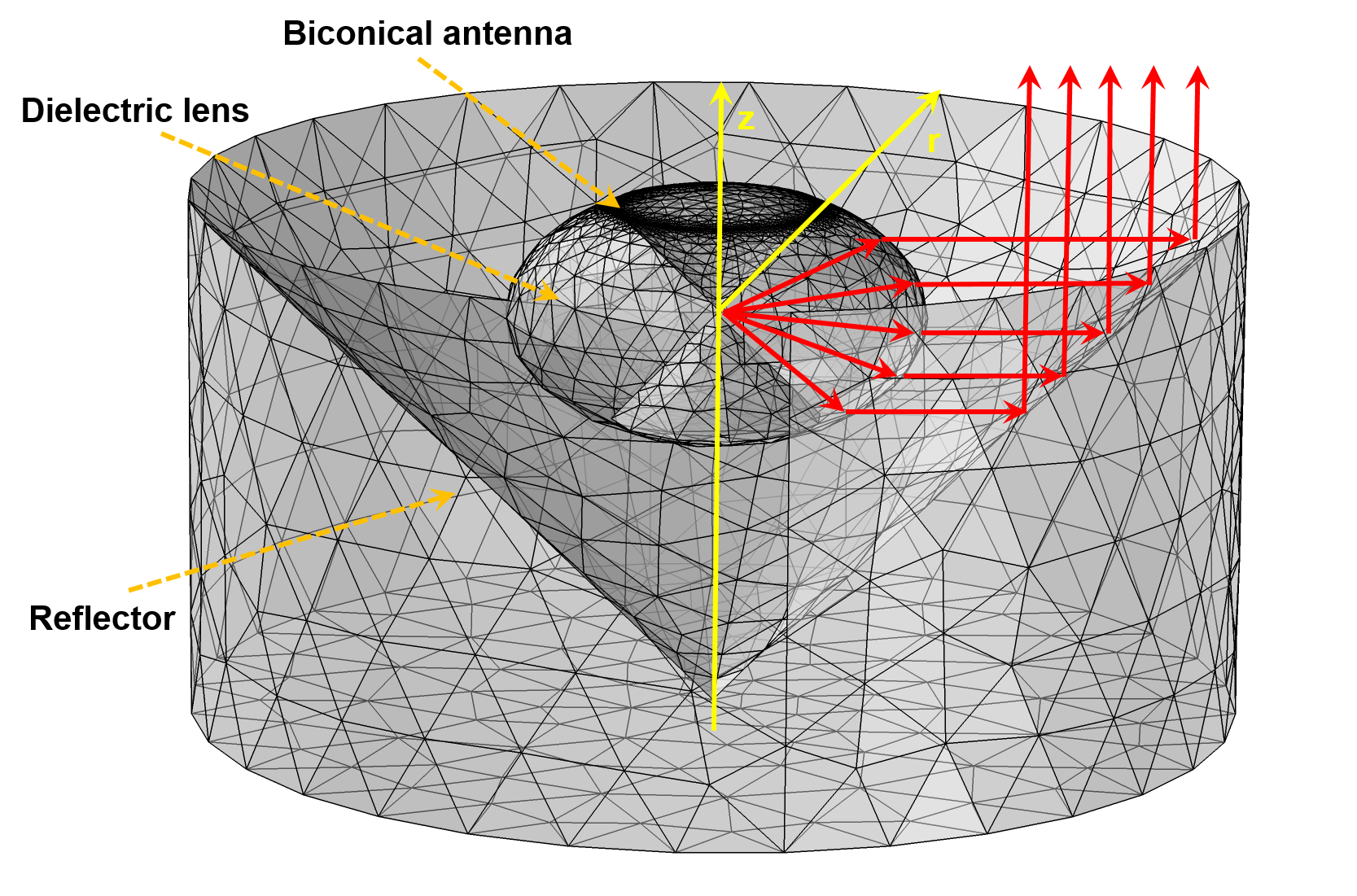}
\caption{\small Scheme showing a rendering of the biconical antenna equipped with
dielectric lens and reflector, as described in the manuscript. Here the emission axis is vertical.
}\label{Fig:figure1}
\end{figure}

The antenna device is made of two conductive cones, having an 
angle of  $\pi /2$ at the vertex, separated by a little gap. These 
conductors are supplied by an RF signal via a coaxial cable passing through 
one of the cones. 
The assembly is embedded in a volume  occupied by a linear homogeneous medium, 
featuring a relative permittivity coefficient $\varepsilon_r>1$,
whereas  the magnetic  permeability coefficient is $\mu_0$. 
The shape of the dielectric is designed in order to fulfill specific refraction properties as
explained in  \cite{chiolerio}. A final conical metal reflector is added to the transmitting device,
so obtaining the complete antenna as shown in Fig. \ref{Fig:figure1}.

\vspace{.5cm}
\vspace{.5cm}

The dream is to get an emitted signal with perfectly parallel rays,
corresponding to infinite directivity.
We will call {\sl emission axis} (z) the 
straight-line naturally associated with the axis of both the cones and the reflector (Fig. \ref{Fig:ruota}).
The electromagnetic wave, during its transition from inside the coax to the free space,
does not break the lines of force of the magnetic field. These remain closed curves encircling 
the emission axis. In this way, at all the steps of the transition, the whole signal maintains 
a cylindrical symmetry. Thus, the fields are zero near the emission axis
and are polarized by following a kind of a circular fashion (Fig. \ref{Fig:ruota}, right).

\vspace{.4cm}
\begin{figure}[h!]
\centering
\includegraphics[width=8.1cm,height=3.7cm]{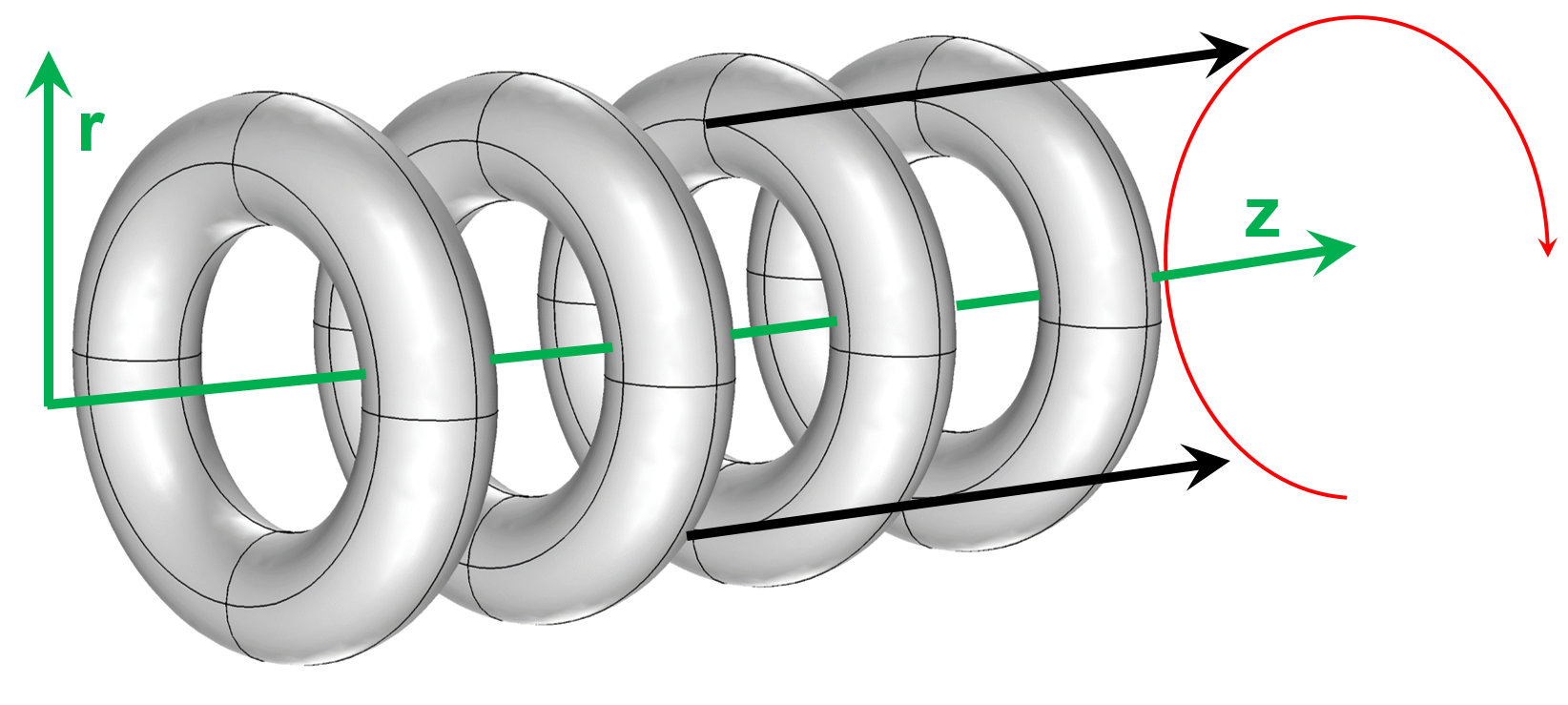}\vspace{-.5cm}
\includegraphics[width=8.1cm,height=3.7cm]{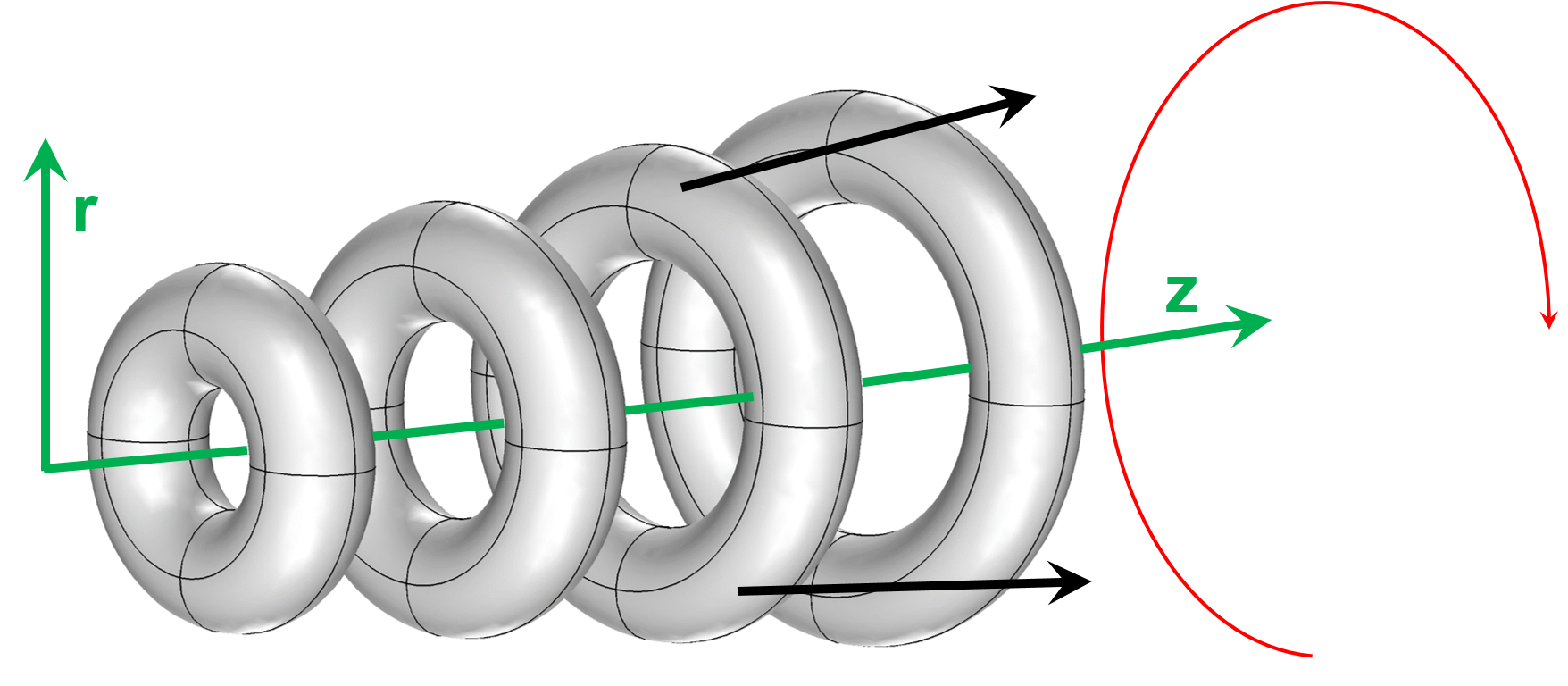}
\caption{\small Behavior of a sinusoidal signal emerging from the antenna:
perfect directivity (top), good directivity (bottom).
Here the emission axis is horizontal. 
}\label{Fig:figure2}
\end{figure}

\vspace{.2cm}
\begin{figure}[h!]
\centerline{\includegraphics[width=5.5cm,height=5.5cm]{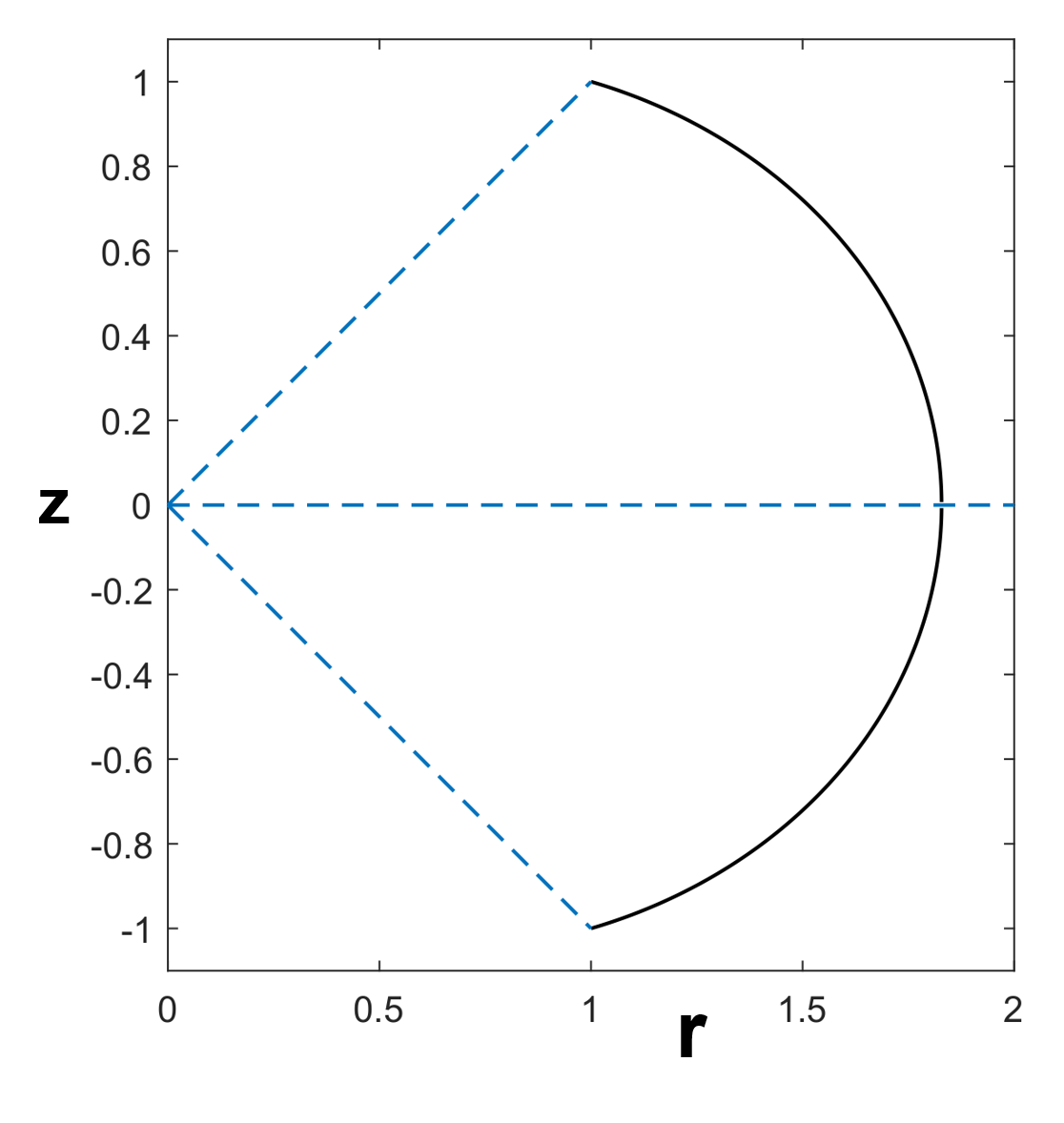}\hspace{.6cm}
\includegraphics[width=6.5cm,height=5.5cm]{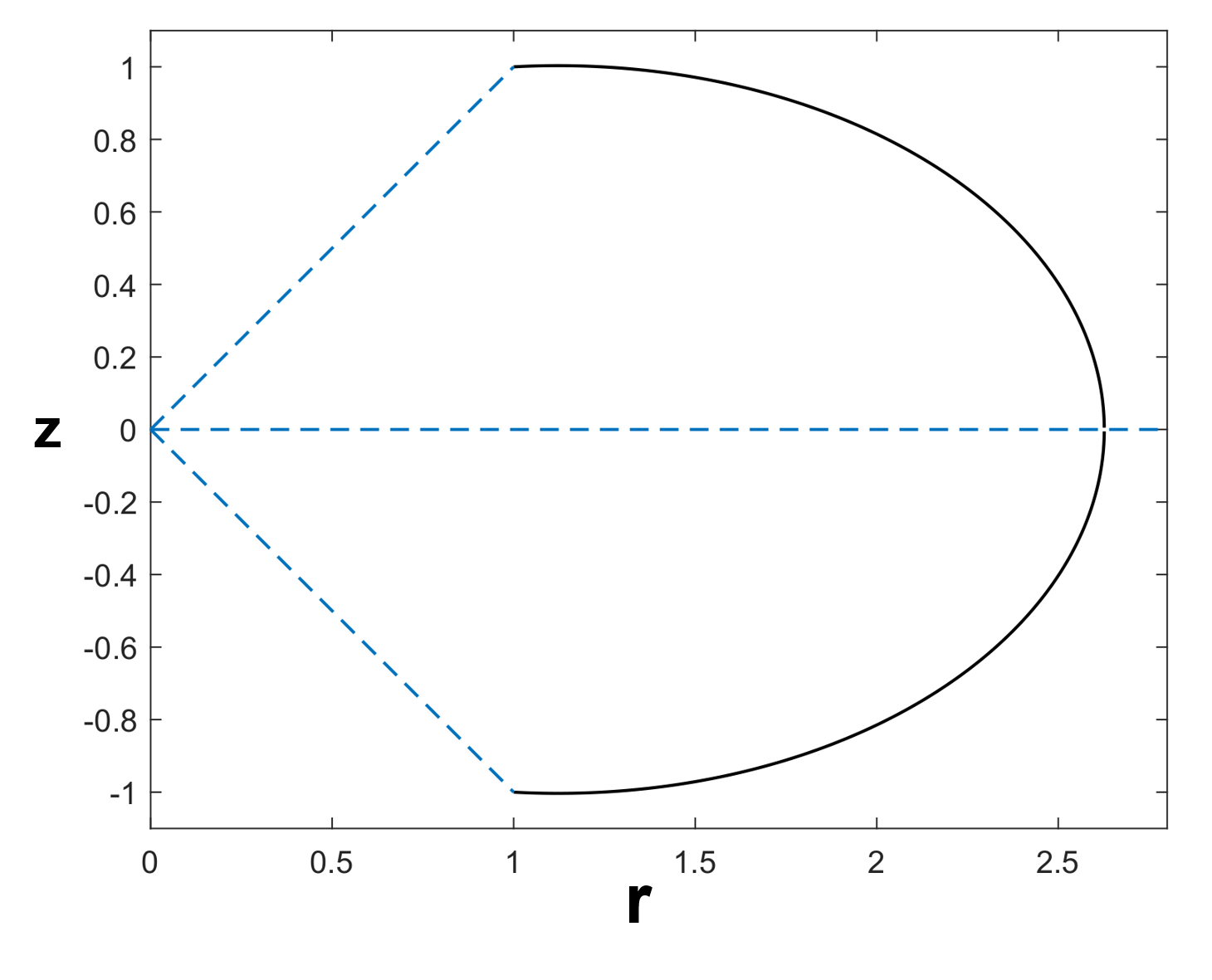}}
\caption{\small Profile of the dielectric lenses: Type A (left), Type B (right). The unit
of measure of the distances is normalized.
 The vertical axis is
parallel to the emission one, the horizontal axis indicates the radius of the device.
}\label{Fig:figure3}
\end{figure}

\vfill

The shape of the emitted wave, modulated by sinusoidal frequency, is expected to look at its best as a
sequence of equal toroidal regions (see Fig. \ref{Fig:figure2}, top), progressing at the speed of
light. Nevertheless, this
would be the ideal behavior. Indeed, in practical experiments, 
 slightly diverging signals are obtained, as for instance in  Fig. \ref{Fig:figure2}, bottom. 
We believe however that there 
is quite enough room to ameliorate these performances (see the discussion in the concluding 
section \ref{sect:options}).

The dielectric lens-equipped antennas that have been tested,
feature a core made by an ensemble of two copper cones and
the whole structure embedded in a dielectric, namely polylactic
acid (PLA), shaped by a 3D printing process by means of a
fused deposition modelling (FDM) machine. We recall that
the electric relative permittivity coefficient for PLA is about
$\epsilon_r \approx 3.5$ \cite{huber}. Each dielectric is composed of two symmetric
parts glued together, with a glue having a similar permittivity constant.
On the other hand, the 3D printing of the entire structure
would have been rather complicated.

\vspace{.3cm}
\begin{figure}[h!]
\centering
\includegraphics[width=5cm,height=4.2cm]{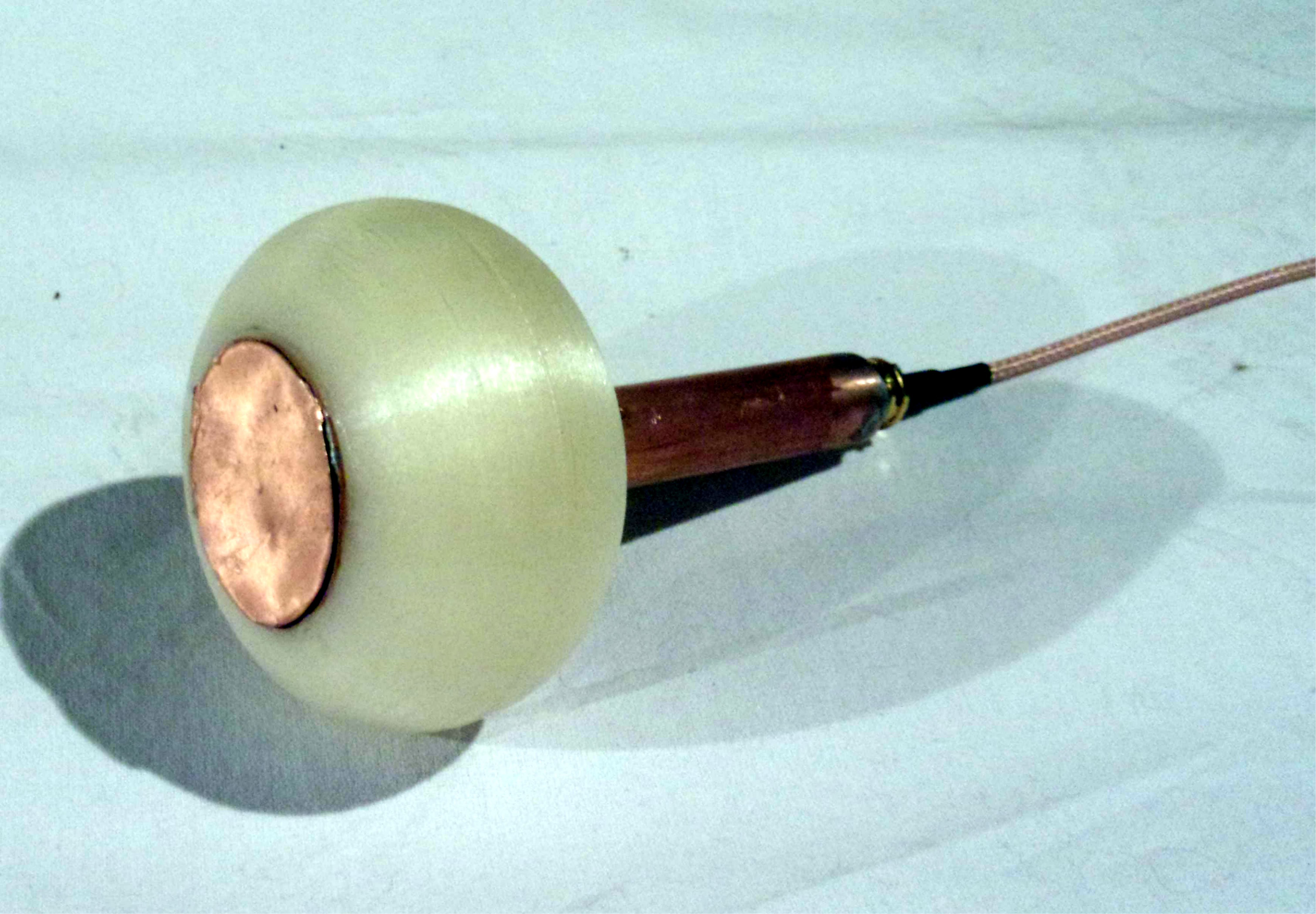}\hspace{.8cm}
\includegraphics[width=6cm,height=4.2cm]{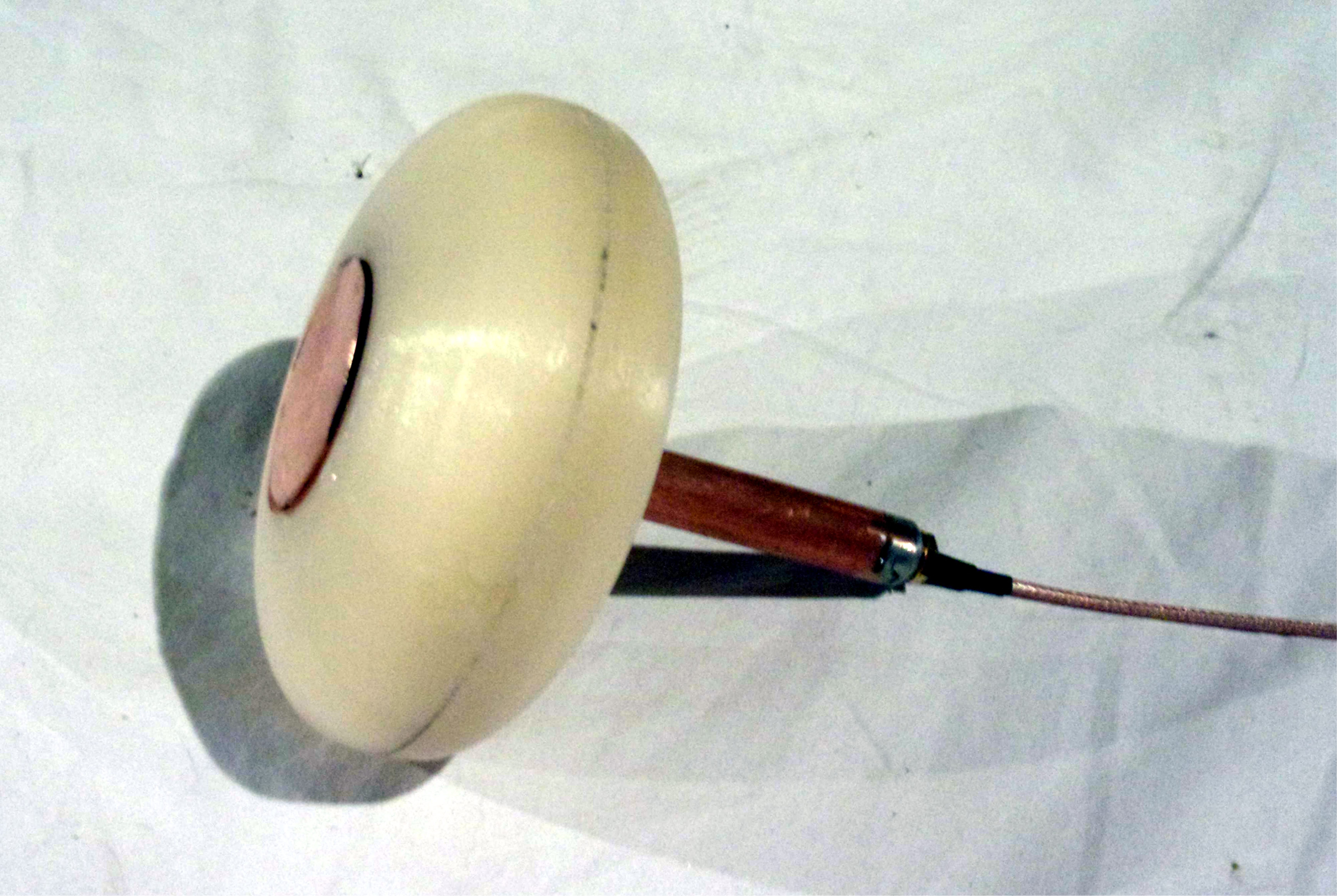}
\caption{\small Biconic components supplied with their 3D printed dielectric: Type A (left), Type B (right).}
\label{Fig:newantennas}
\end{figure}
\vspace{.2cm}

The antenna of Type A features two copper cones of base
3 cm, height 1.5 cm. The dielectric part has a width of
approximately 5.4 cm and height of 3 cm. Its shape has been
chosen as suggested in [16], section \ref{sect:options}, by a proper matching
of the value of the relative permittivity, chosen to be $\epsilon_r =4$ 
(thus a bit larger than the coefficient corresponding to PLA).
The dielectric profile can be seen in Fig. 3 (left), and the final
realization in Fig. 4 (left)

Type B features two copper cones having base 4 cm, height
2 cm. The dielectric lens has a width of approximately 10.4
cm and height of 4 cm. This time the profile has been chosen
to a corresponding value of $\epsilon_r = 1.8$ (approximately a half
of that of PLA). The dielectric profile can be seen in Fig. 3
(right), and the final realization in Fig. 4 (right).
Even though the permittivity coefficients of both antenna
prototypes do not match with those theoretically recommended, the 
choices are justified by the willing to test two
rather different conditions. Since the 3D printing process for
the dielectrics is quite time consuming, a very limited number
of prototypes at this preliminary stage was produced.

\section{Tests in anechoic chamber}\label{sect:tests}

\begin{figure}[h!]
   \centering{ \includegraphics[width=12.cm,height=5.8cm]{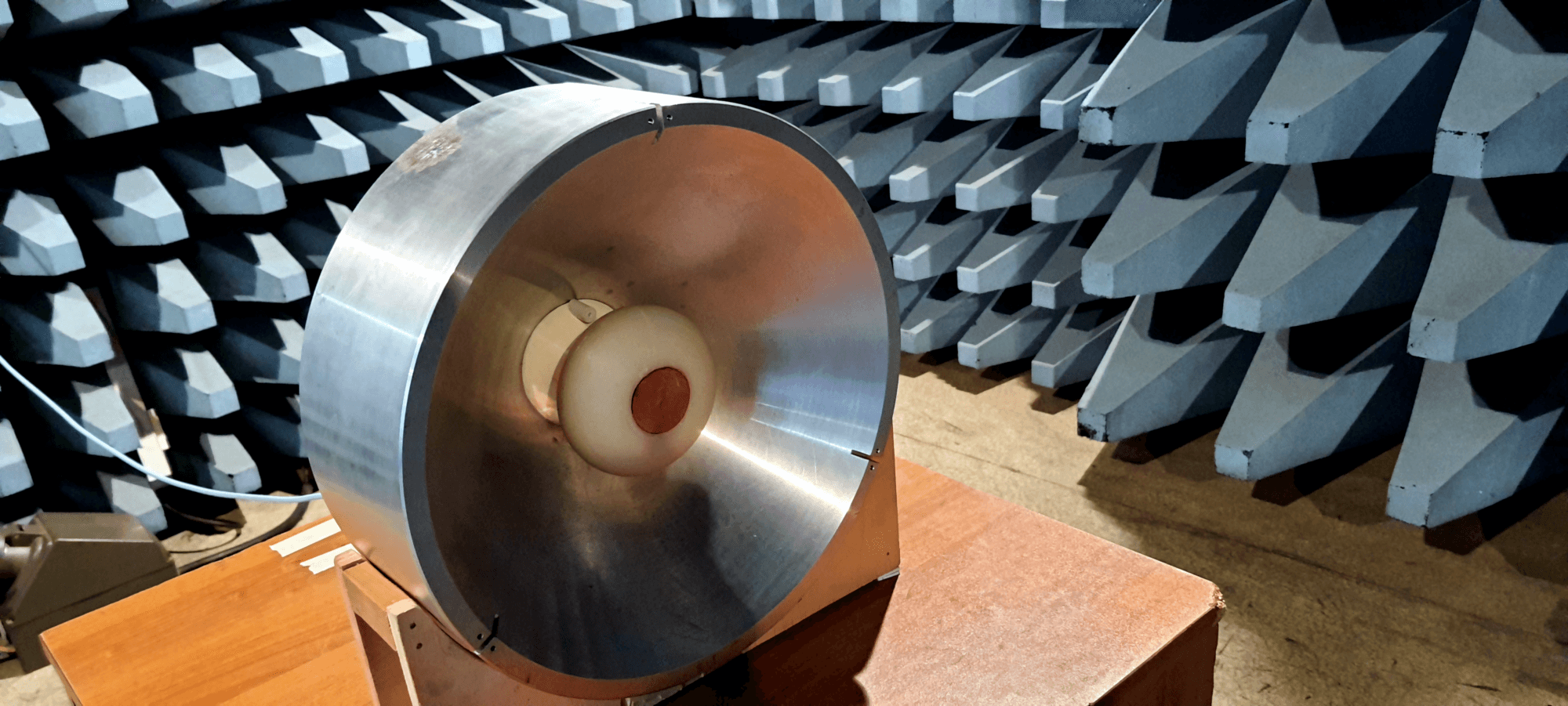}}
    \hphantom{}\vspace{.3cm}
  \centering{\includegraphics[width=12.cm,height=5.8cm]{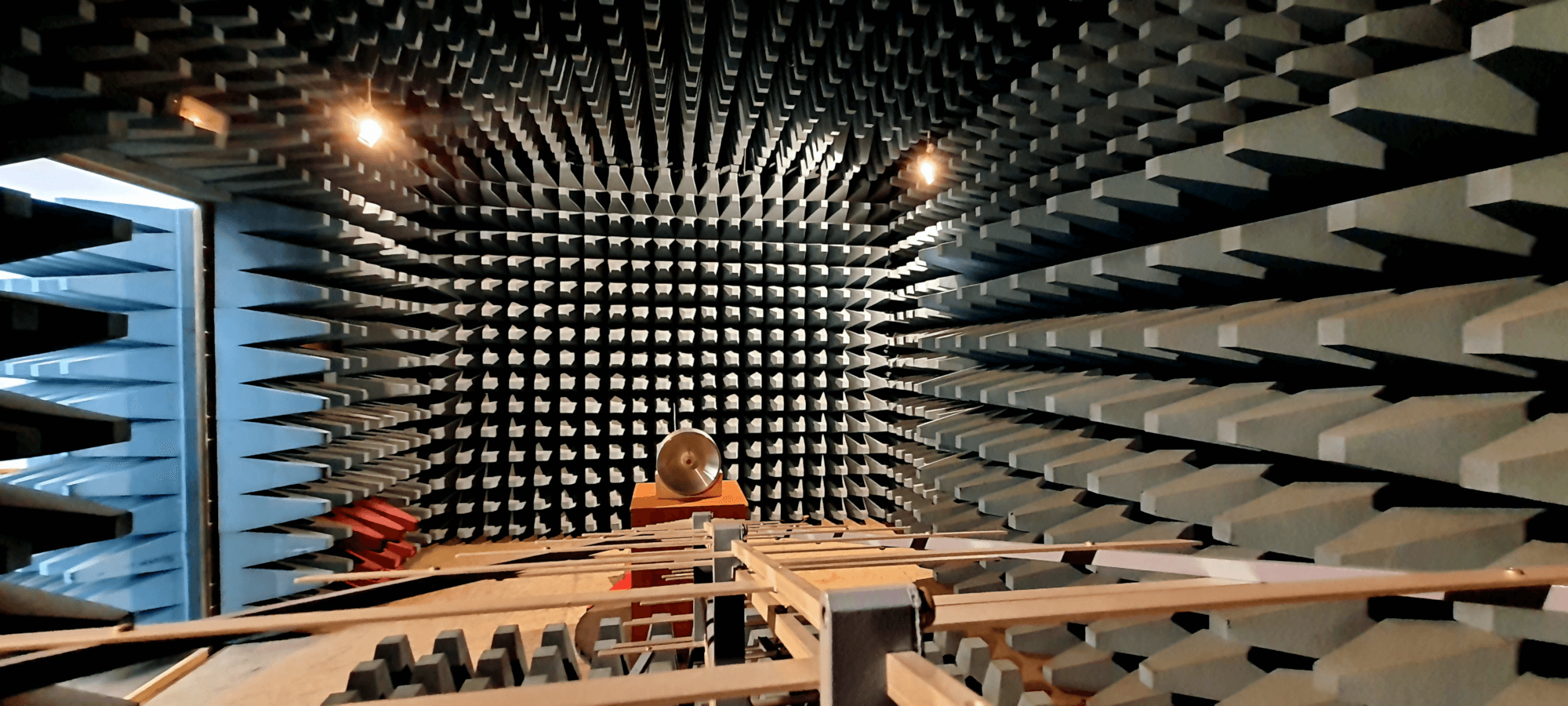}}
    \hphantom{}\vspace{.3cm}
    \caption{\small Layout of the measurement area.
    Detail showing the aluminum reflector (top), hosting the biconical antenna surrounded by 3D printed dielectric.
    Overview of the anechoic chamber (bottom), where the log-periodic receiver can be glimpsed.
    }
    \label{fig:anecoic}
\end{figure}

\begin{figure}[h!]
\centering{
{\includegraphics[width=5.2cm,height=5.6cm]{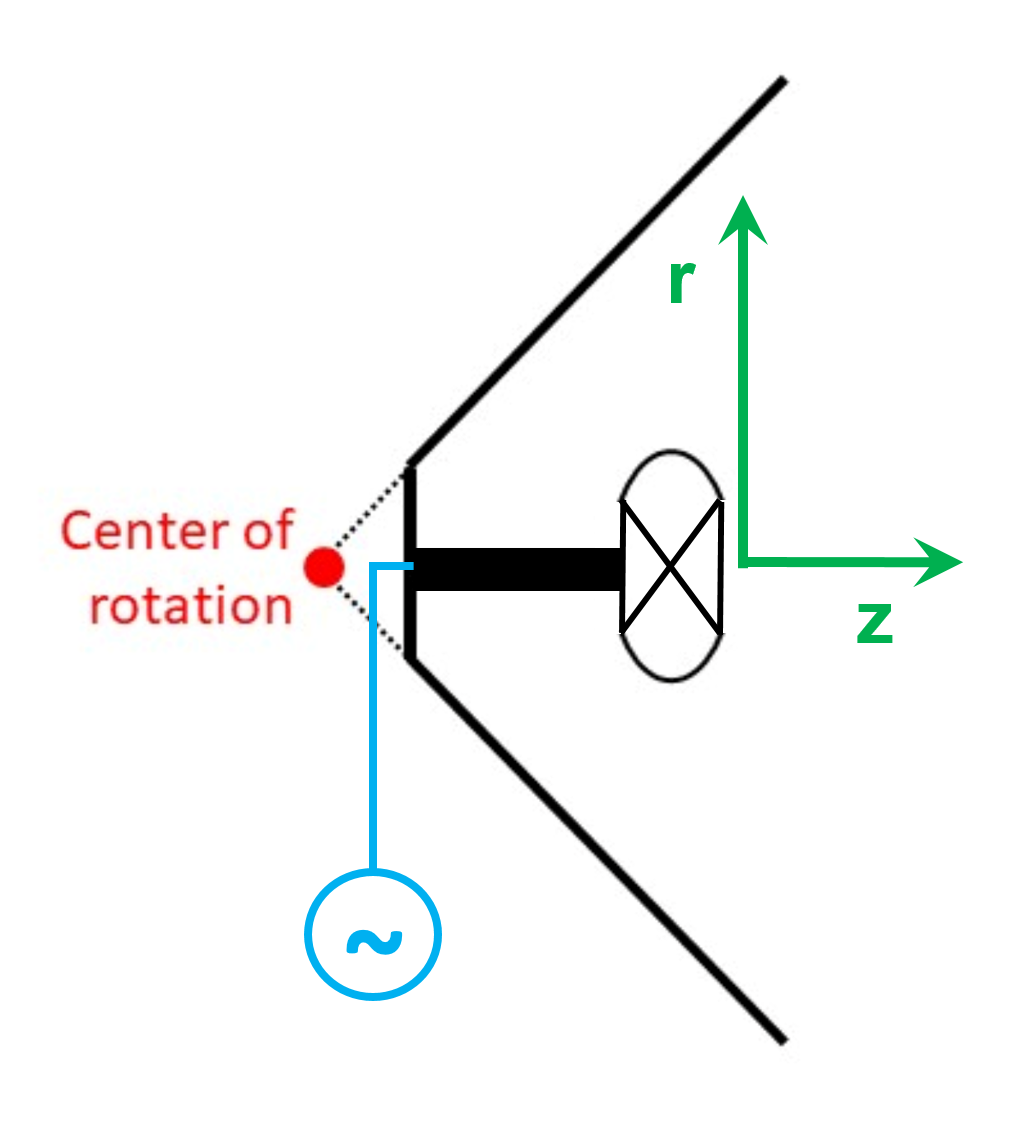}}\hspace{.5cm}
{\includegraphics[width=6.1cm,height=5.7cm]{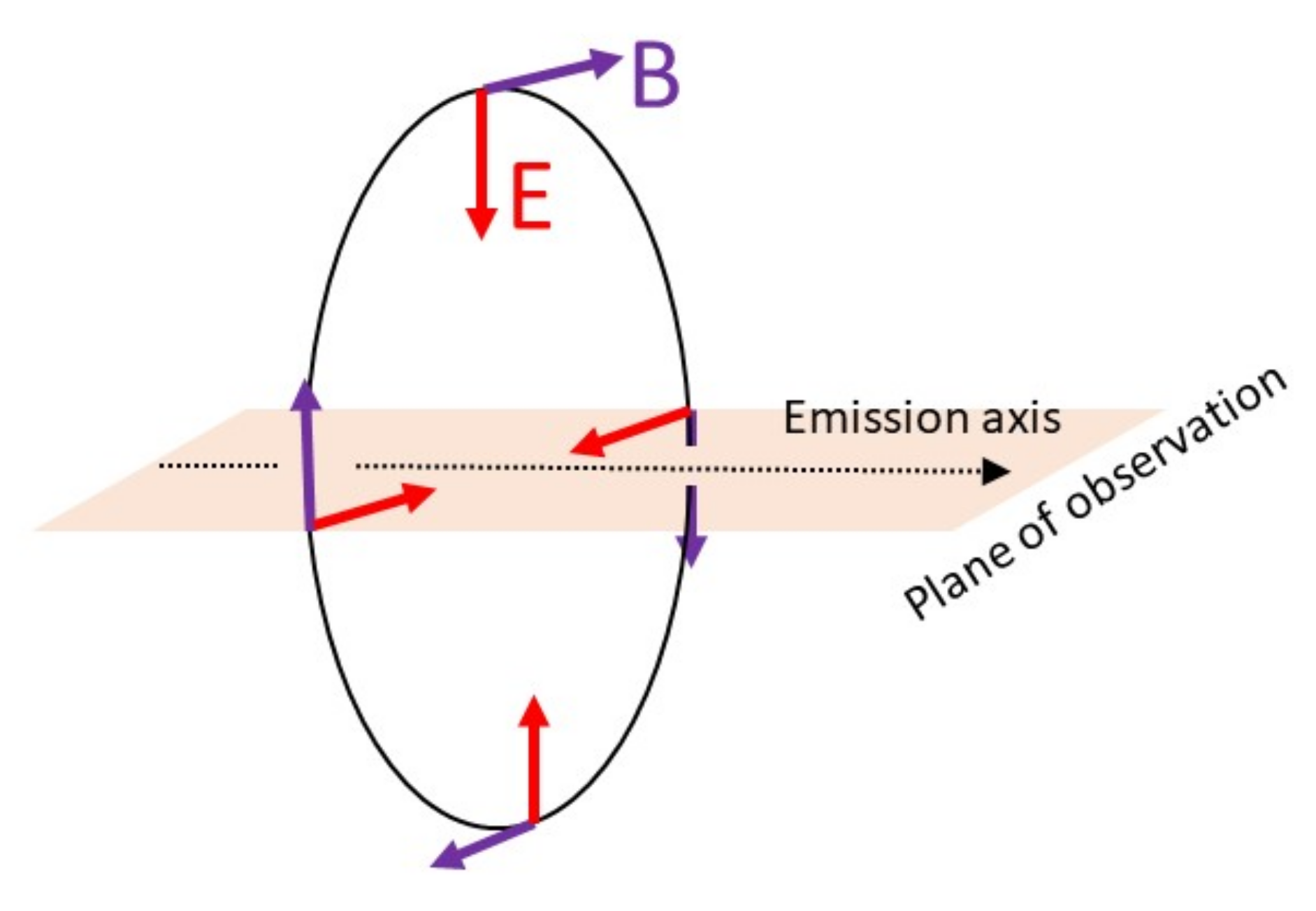}}
}
\caption{\small Fulcrum of rotation of the antenna (left), as seen from above, depicted with circuital connection to the waveform generator and polar axes. The expected electric signal, measured on the horizontal plane containing the emission axis, is schematically displayed on the right. Note that the electric field is radial and the magnetic field rotates around the emission
axis, which is parallel to the Poynting's vectors. Both fields vanish at the
center of the circle.}\label{Fig:ruota}
\end{figure}

A conic reflector (see Figs. \ref{Fig:figure1},  \ref{fig:anecoic},  \ref{Fig:ruota}) 
is used to convey the  radiated waves towards the
receiver. It features a 32 cm opening diameter, a vertex angle equal to $\pi/2$, and is manufactured using
a lathe from a single piece of aluminium. Each antenna is centered and hosted inside the 
reflector, and finally supplied from the rear side via a SMA connector and cable.

\vspace{.5cm}

The measurements have been performed in an anechoic
chamber equipped with a rotating table, using the following
instrumentation: the tracking generator of a Rohde \& Schwarz
ESBI receiver with an output of 1 mW, a log-periodic Rohde \& Schwarz HL562 (in the frequency range between 500 MHz
and 3 GHz), an Amplifier Research 30S1G3 stage supplying
30 W (in the frequency range between 1 GHz and 5 GHz),
coupled with a horn antenna Rohde \& Schwarz HF906. The
height of the center of the device from the ground is 1 meter
and the distance from the receiver is about 4 meters.

The entire
biconical antenna is rotated horizontally about the vertex of
the reflecting cone as shown in Fig. 6 (left). The angles range
between $0^\circ$ (emission axis pointing towards the receiver) up to
$50^\circ$ at steps of about $2^\circ$. For symmetry reasons, the tests are
independent of the rotation sense, so that only one side was
exploited. Figure 6 (right) gives an idea of the polarization of
the electric field on the plane of observation.

\vspace{.6cm}

\begin{figure}[h!]
\centering{
   {\includegraphics[scale=.24]{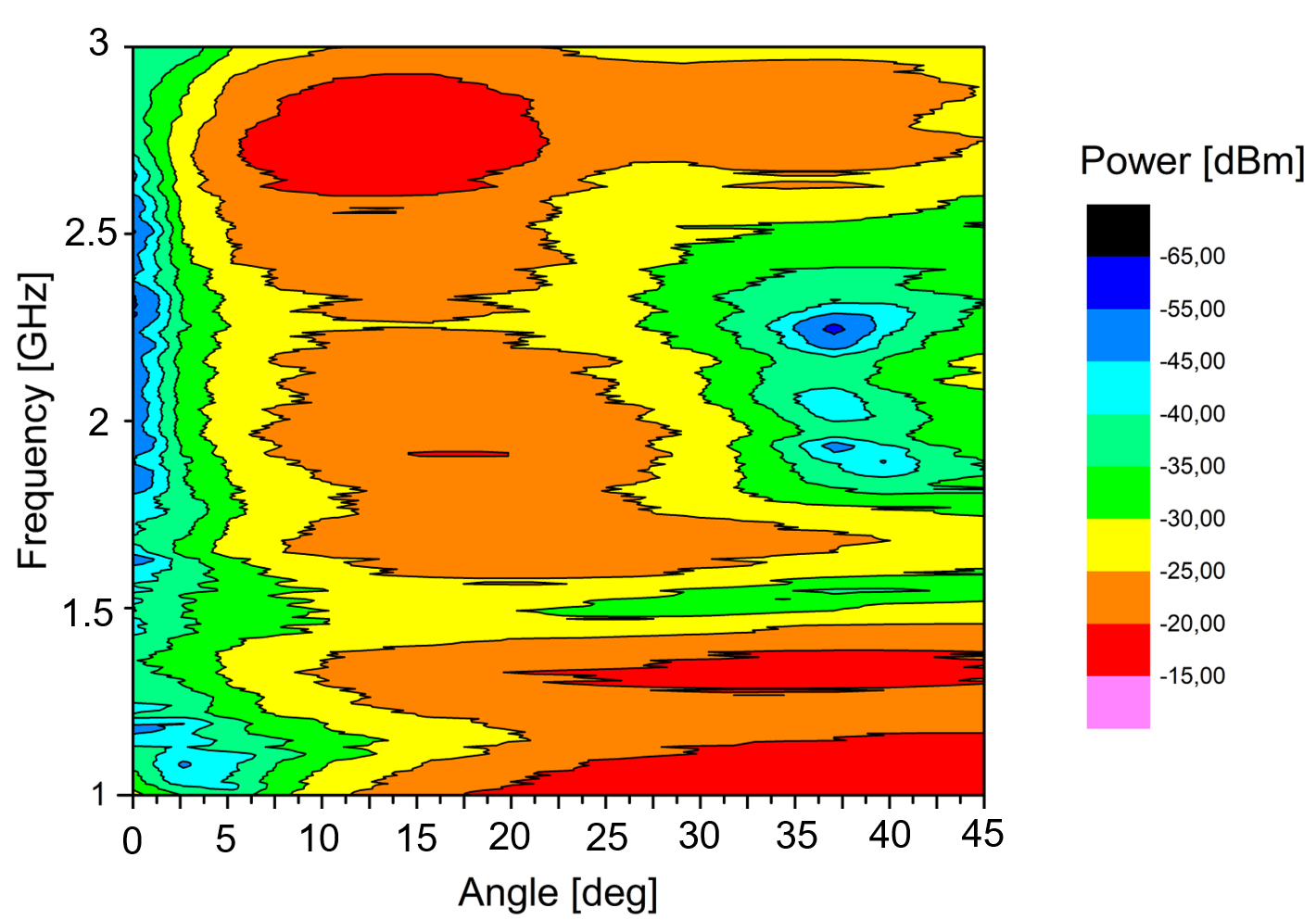}}\hspace{.2cm}
   {\includegraphics[scale=.24]{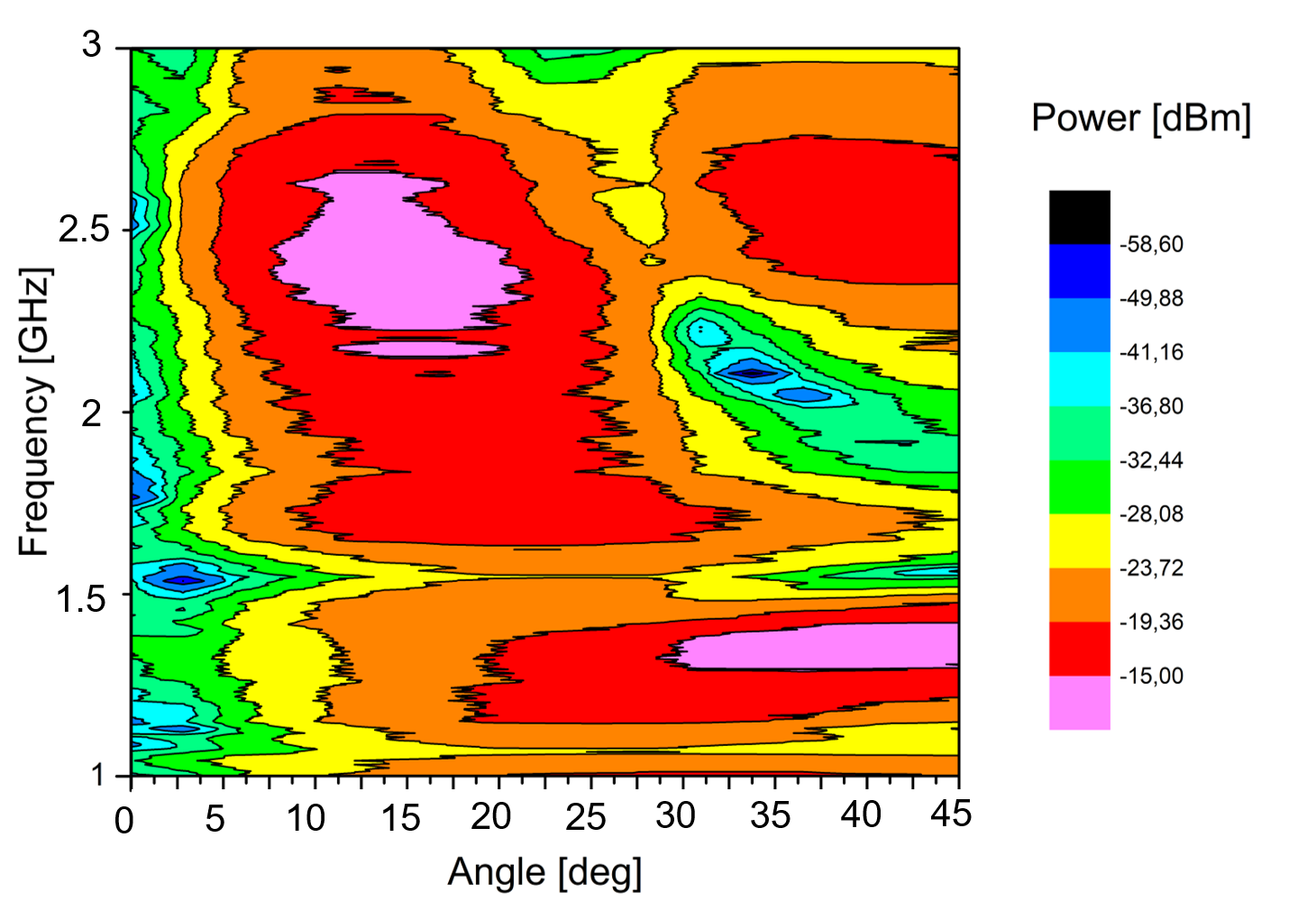}}
  }
  \vspace{.2cm}
    \caption{\small 2D plots of the antenna emission power in dBm, recorded with a log-periodic 
    antenna: Type A (left), Type B (right). The frequency ranges from 1 GHz up to 3 GHz.
    }
    \label{fig:2Dmaps}
\end{figure}

Figure \ref{fig:2Dmaps} shows the 2-dimensional contour plots of the measured received power,
with angles from $0^\circ$  to $45^\circ$, and  frequencies ranging between 1 GHz and 3 GHz.
The unity of measure of the instruments is given in Decibels per mW (dBm). 
As the scale is logarithmic with argument less than 1, the greatest emission strengths are 
those close to zero. 

We are not aware of the impedance properties of the transmitting antennas, therefore we are
unable to judge the efficiency of our devices. We believe that this is irrelevant at this stage, since here the primary concern
is the direction of propagation of the signal.
We have relatively good directivity when, for a given frequency, the patterns tend
to be concentrated towards the small angles.
For example,  both antennas show scarce directivity below 1.5 GHz or above 2.5 GHz, while 
the intermediate frequency range looks much better. For frequencies above 2.5 GHz 
a secondary lobe appears. We will provide more explanations of these results in the concluding section \ref{sect:options}.

\vspace{.3cm}
\vspace{.5cm}

\begin{figure}[h!]
\centering{
   {\includegraphics[scale=.22]{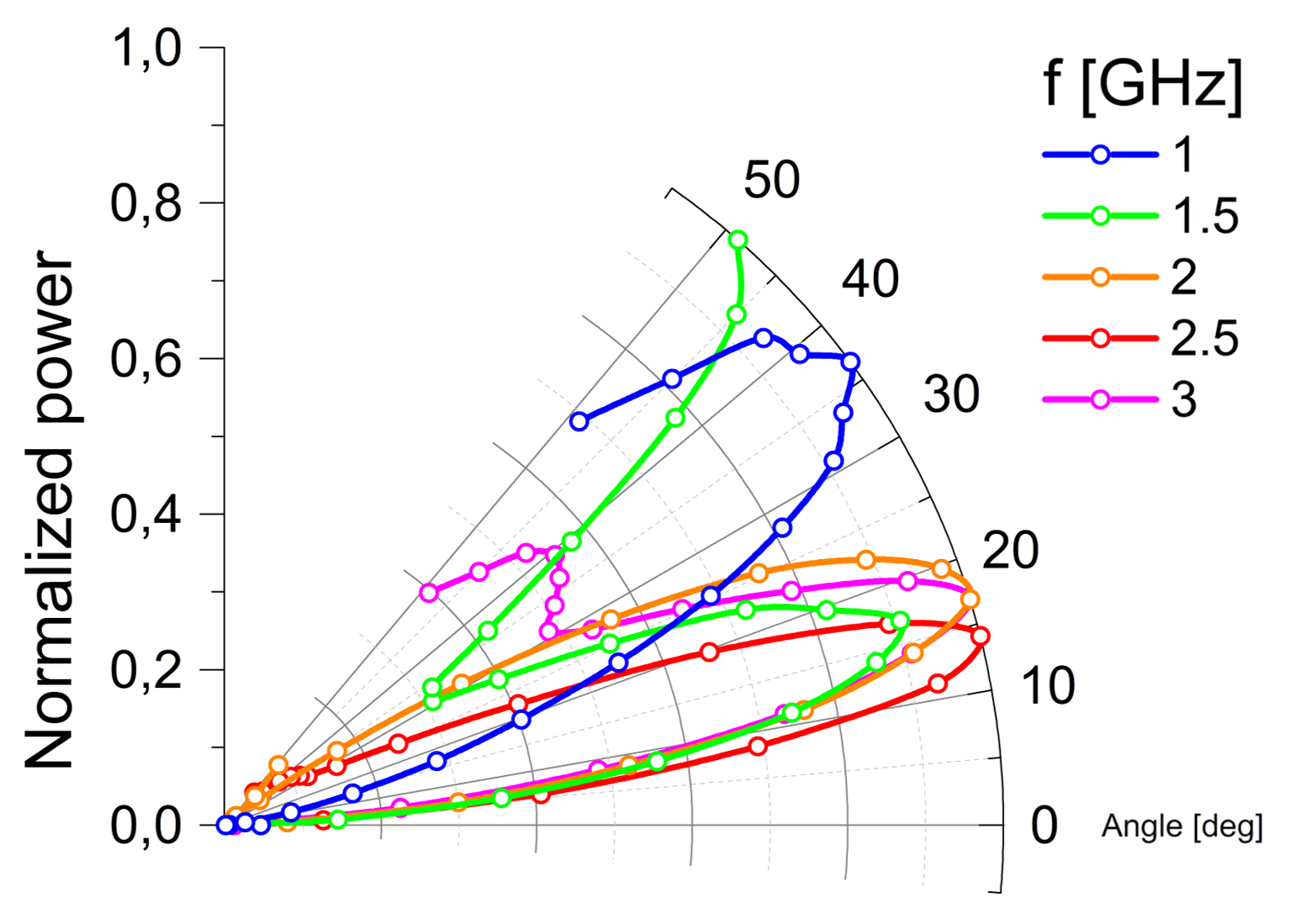}}\hspace{.2cm}
  { \includegraphics[scale=.22]{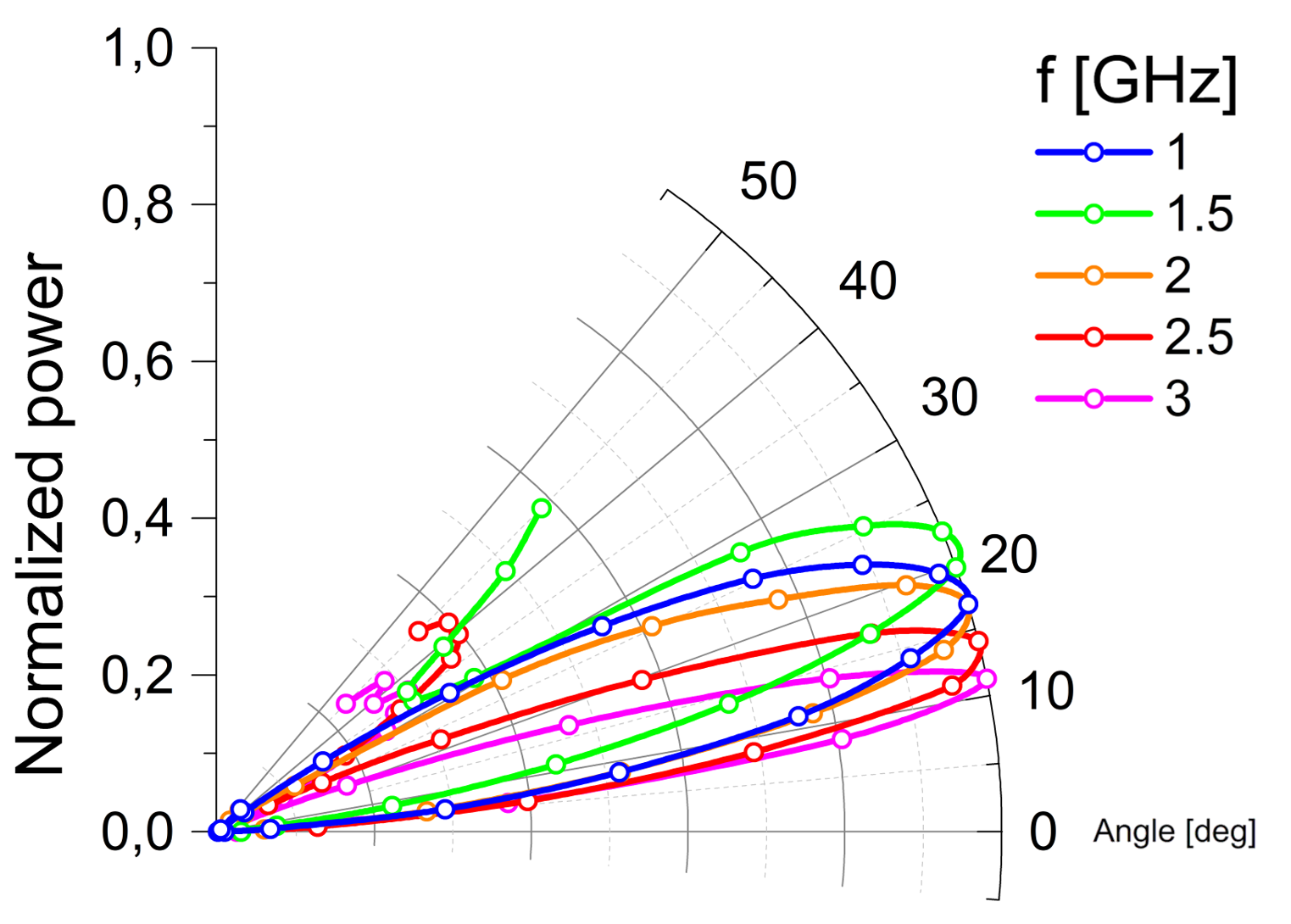}}
  }
    \caption{\small Polar diagrams of the normalized emission powers at selected frequencies 
    1, 1.5, 2, 2.5 and 3 GHz, recorded with the log-periodic receiver: Type A (left), Type B (right).
    }
    \label{fig:polarmaps}
\end{figure}

A more practical visualization is represented by the polar charts of  Fig. \ref{fig:polarmaps}. 
Here, we switched from dBm to the power expressed in mW using the conversion formula: 
$P=10^\wedge \{{\rm dBm}/10^{-3}\}$.

\vspace{.5cm}
\vspace{.5cm}

The results relative to the antenna of
Type B show a more robust diagram of the emission pattern, where the principal lobes are
comprised between $10^\circ$ and $20^\circ$. Antenna 
of Type A shows the principal lobes distributed over a broader range, comprised between 
$10^\circ$ and $50^\circ$. In addition, there are secondary lobes, showing up at
frequencies above 1.5 GHz. These last are comprised in the range between $30^\circ$ and $50^\circ$.
The optimal directivity was found for a frequency of about
2.3 GHz (antenna of Type A), with a lobe  centered around $14^\circ$ featuring a FWHM of $16.1^\circ$.
Regarding the antenna of Type B, we get the best results when the frequency is about $1.69$ GHz
and the lobe attains its maximum around $19.6^\circ$, with a FWHM of $20.7^\circ$.

Theoretically, the adoption of a dielectric shape where the value $\varepsilon_r >3.5$ (we
recall that 3.5 is the relative permittivity of PLA) is overestimated as in Fig. \ref{Fig:figure3}
(left), tends to spread the signal. Conversely, when $\varepsilon_r$ is less than the suggested value 3.5,
as in Fig. \ref{Fig:figure3} (right), the signal tends to be over focussed, with  the consequence that
the rays may cross each other before reaching the reflector, resulting again in a spread of 
the emission. The right balance would follow from testing with many different shapes. This
is however not an option in this preliminary analysis.


\vspace{.3cm}
\begin{figure}[h!]
\centering{
   {\includegraphics[scale=.24]{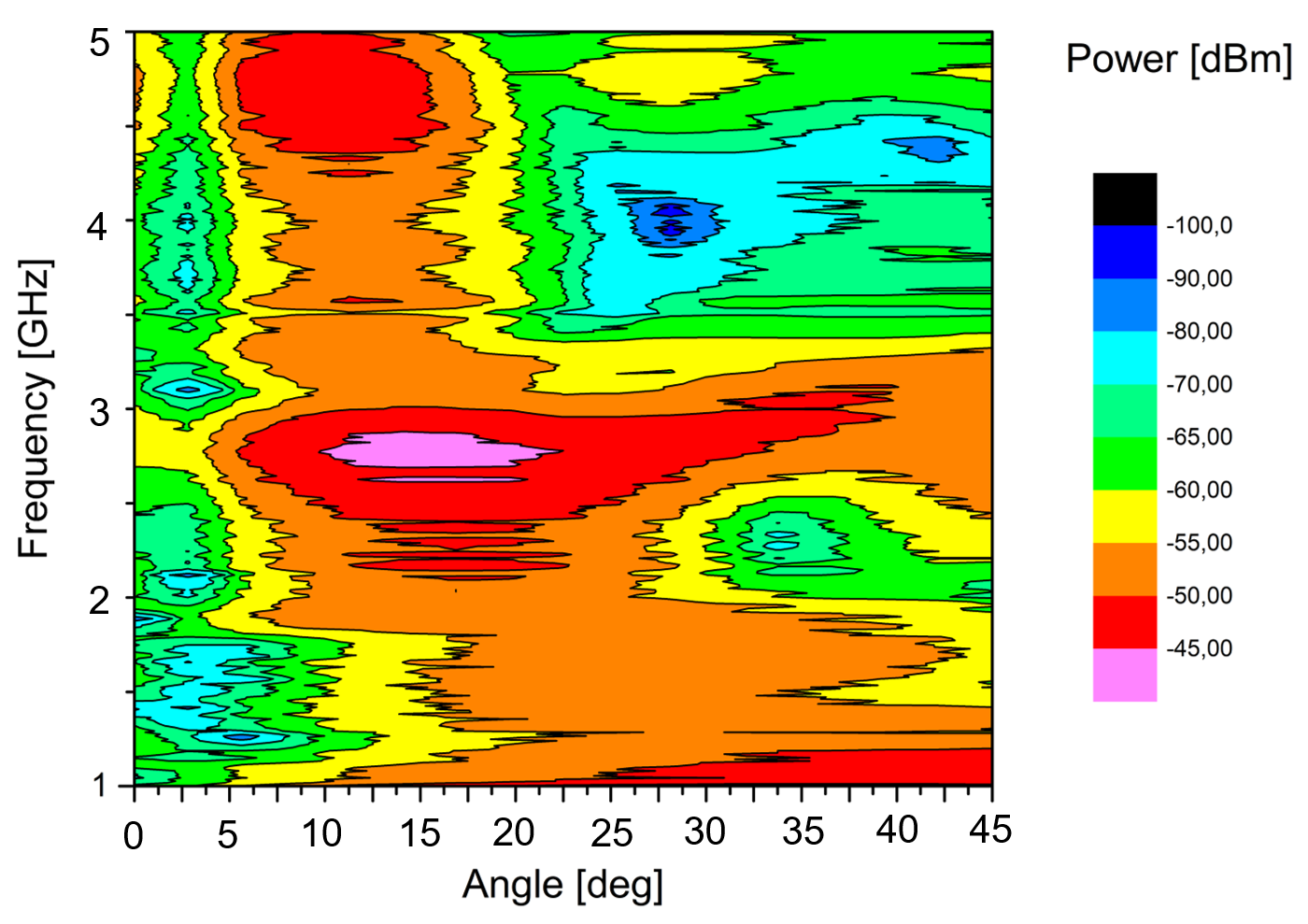}}\hspace{.2cm}
  { \includegraphics[scale=.24]{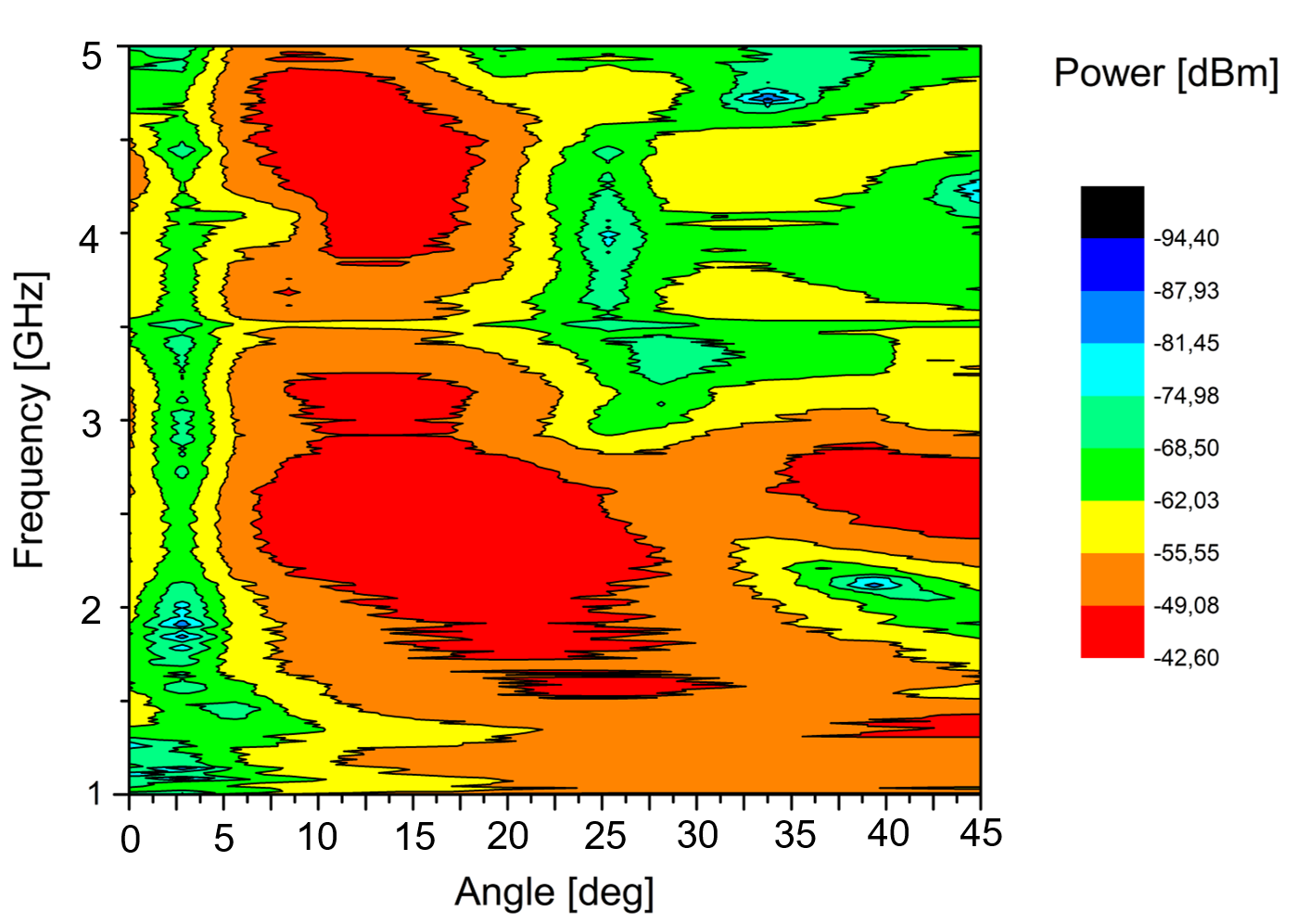}}
    }
    \vspace{.2cm}
    \caption{\small 2D plots of the antenna emission power in dBm, recorded with a horn antenna: 
    Type A (left), Type B (right). The frequency ranges from 1 GHz up to 5 GHz.
    }
    \label{fig:2DmapsHF}
\end{figure}
\vspace{.2cm}

\vspace{.5cm}
\begin{figure}[h!]
\centering{
   {\includegraphics[trim = 1.9cm 1.3cm .3cm .4cm, clip, scale=.21]{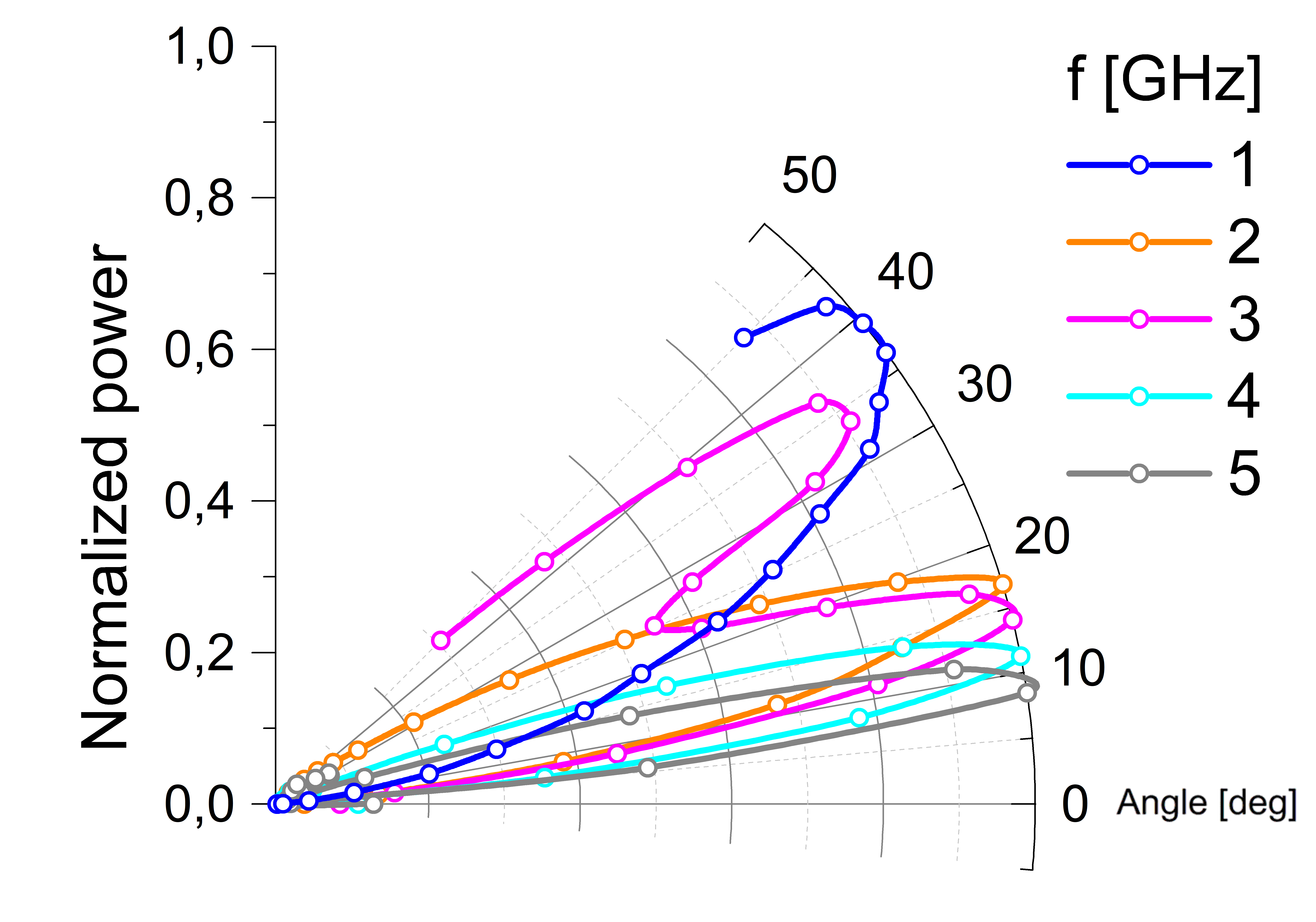}}\hspace{.5cm}
  { \includegraphics[trim = 1.9cm 1.3cm .3cm .4cm, clip, scale=.21]{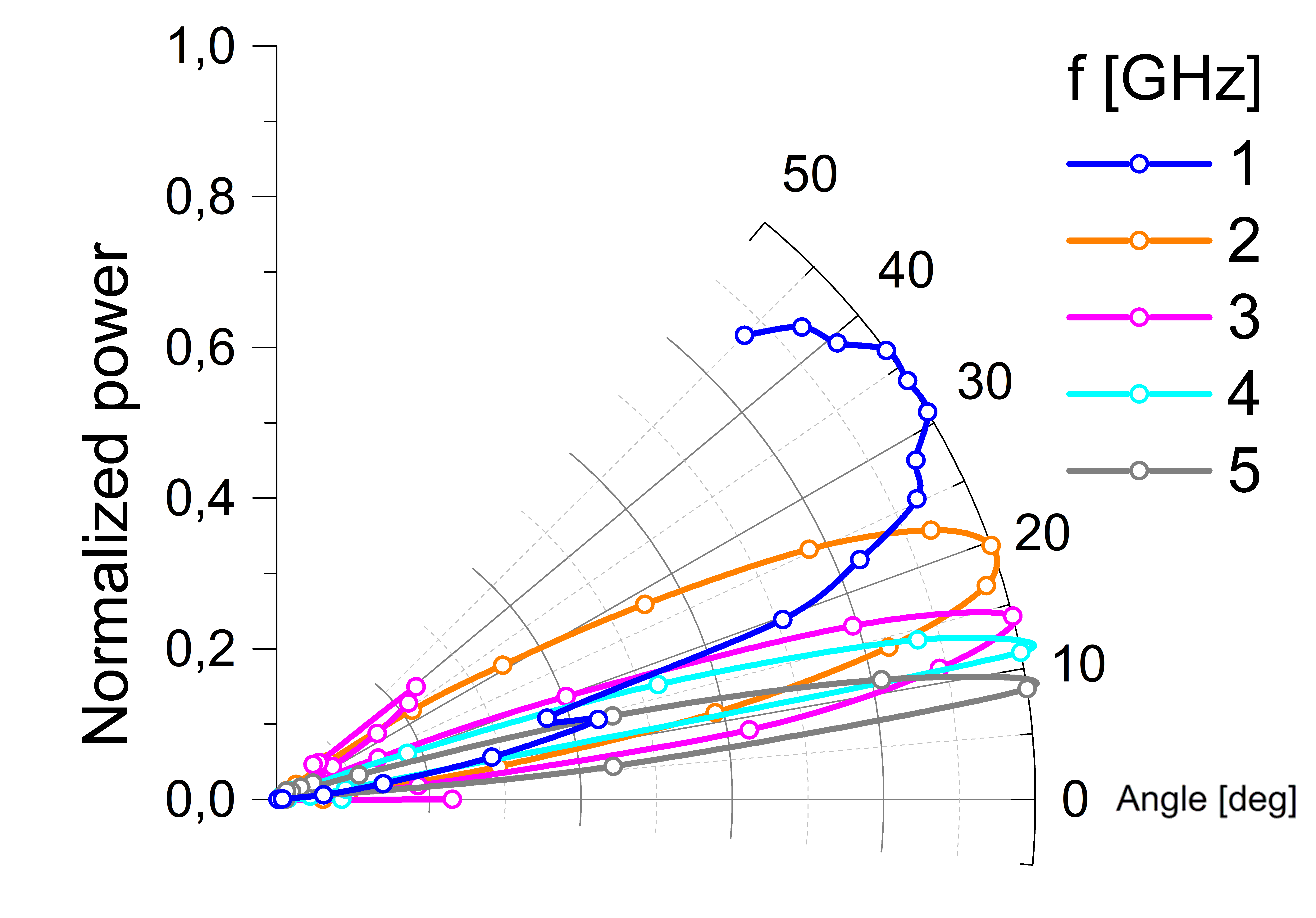}}
    }
		\vspace{-.3cm}
    \caption{\small Polar diagrams of the normalized emission powers at selected frequencies 
    1, 2, 3, 4 and 5 GHz, recorded with the horn receiver: Type A (left), Type B (right).
    }
    \label{fig:polarmapsHF}
\end{figure}
\vspace{.1cm}

Further data are available. These were recorded with a horn type antenna with an extended  frequency
ranging between 1 GHz and 5 GHz.
Figure \ref{fig:2DmapsHF} shows the 2-dimensional contour plots of the emission power, measured for
angles between $0^\circ$ and $45^\circ$.
Both antennas display a broad emission below 3.5 GHz, while in the higher frequency range 
the directivity looks extremely good. 

The same results are displayed in the polar plane in Fig. \ref{fig:polarmapsHF}, for some
selected frequencies. These measurements confirm that
the antenna of Type B provides better emission patterns, having in some cases the principal
lobe distributed around $10^\circ$. Antenna of Type A shows principal
lobes distributed over a broader range, comprised between $10^\circ$ and $40^\circ$. 
A prominent secondary lobe
appears for a frequency of 3.5 GHz, and is comprised in the range between $30^\circ$ and $40^\circ$.

The optimal directivity for the antenna of Type A was found for a frequency of about 4 GHz. The corresponding lobe is centered around $8.4^\circ$, featuring a FWHM of $6.4^\circ$.
Concerning the Type B antenna, the best performances are obtained for a frequency of about 3.66 GHz. The corresponding lobe is centered around $11.3^\circ$, featuring a FWHM of $9.9^\circ$.

A first explanation of the better performances at higher frequencies is that both the 
emitting devices are relatively small. The qualitative difference of the outcomes of
figures \ref{fig:polarmaps} and  \ref{fig:polarmapsHF} in the range of frequencies
at the intersection (i.e.: 1 GHz to 3 GHz) can be justified by the type of antenna used
as a receiver. In particular, the log-periodic antenna is much bulkier than the horn antenna.
This  affects  experiments, since, theoretically, the received should
be placed at infinite distance in order to probe high directivity emission patterns.
More insight on these issues will be given in section \ref{sect:options}.

\section{Bare biconical antennas}\label{sect:earlier}

\begin{figure}[h!]
\centering
\includegraphics[width=7.1cm,height=3.7cm]{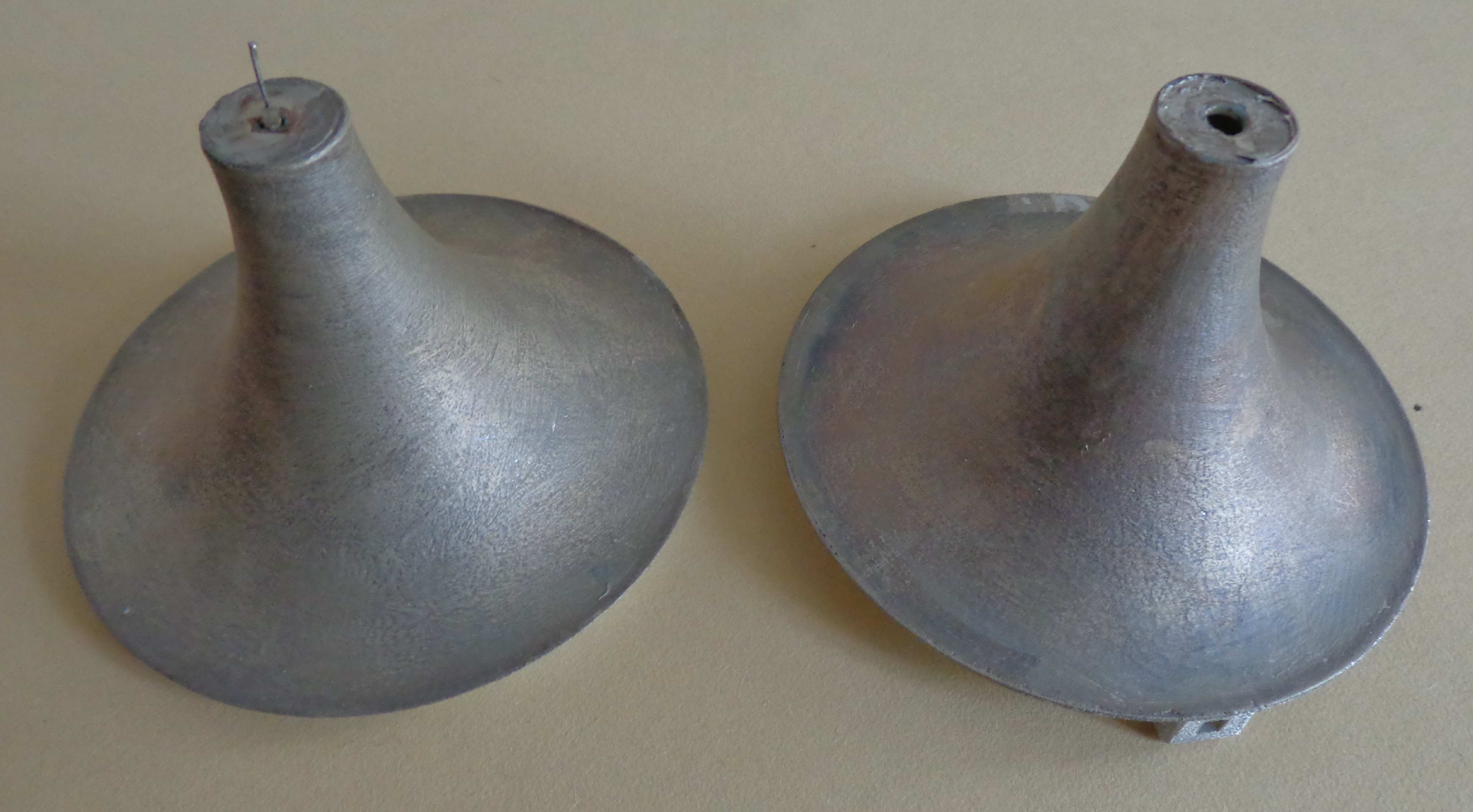}\hspace{.7cm}
\includegraphics[width=3.1cm,height=3.7cm]{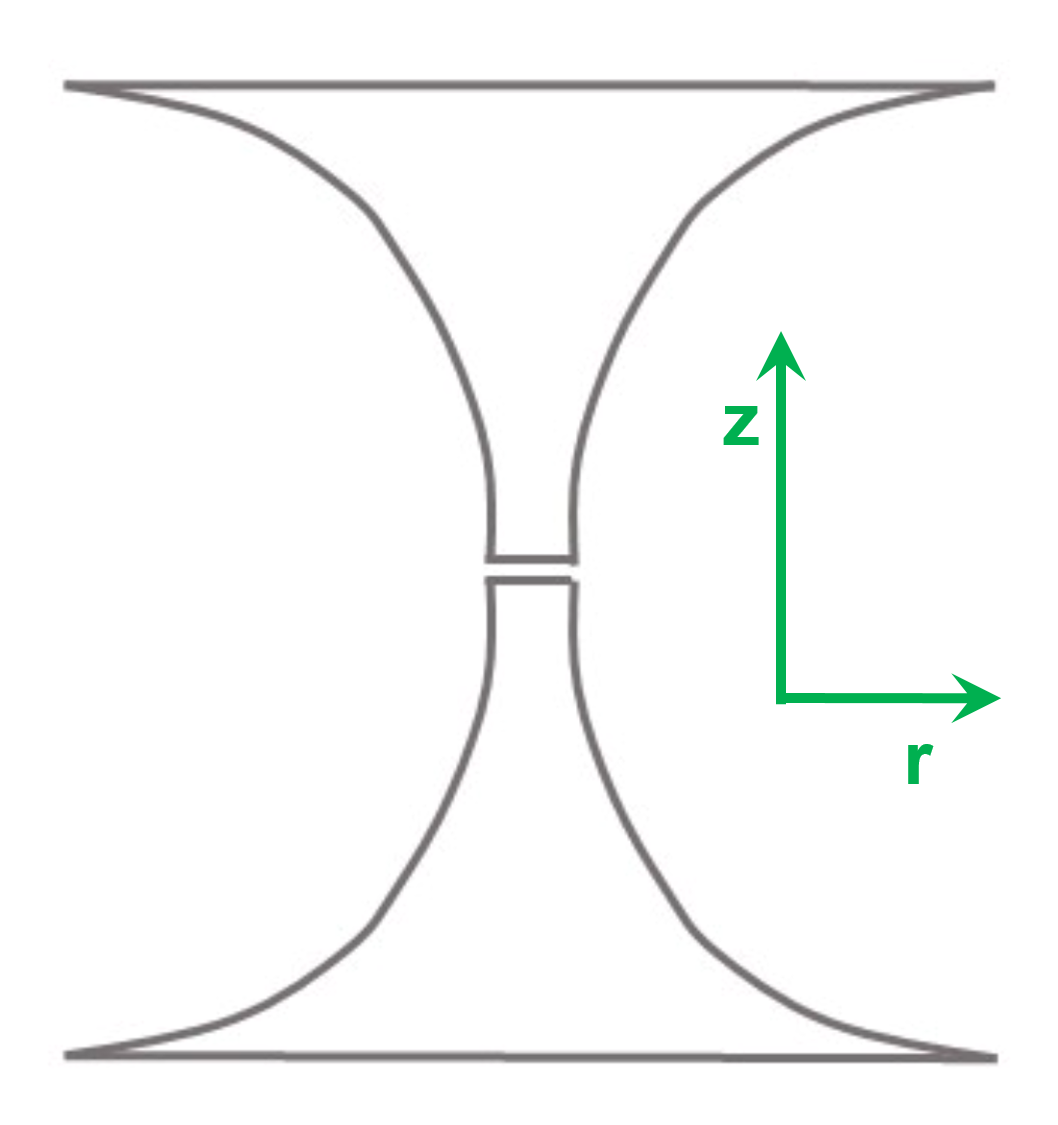}
\vspace{.2cm}
\caption{\small Type C bare biconical antenna.}\label{Fig:cv}
\end{figure}

\begin{figure}[h!]
\centering
\includegraphics[width=7.1cm,height=3.7cm]{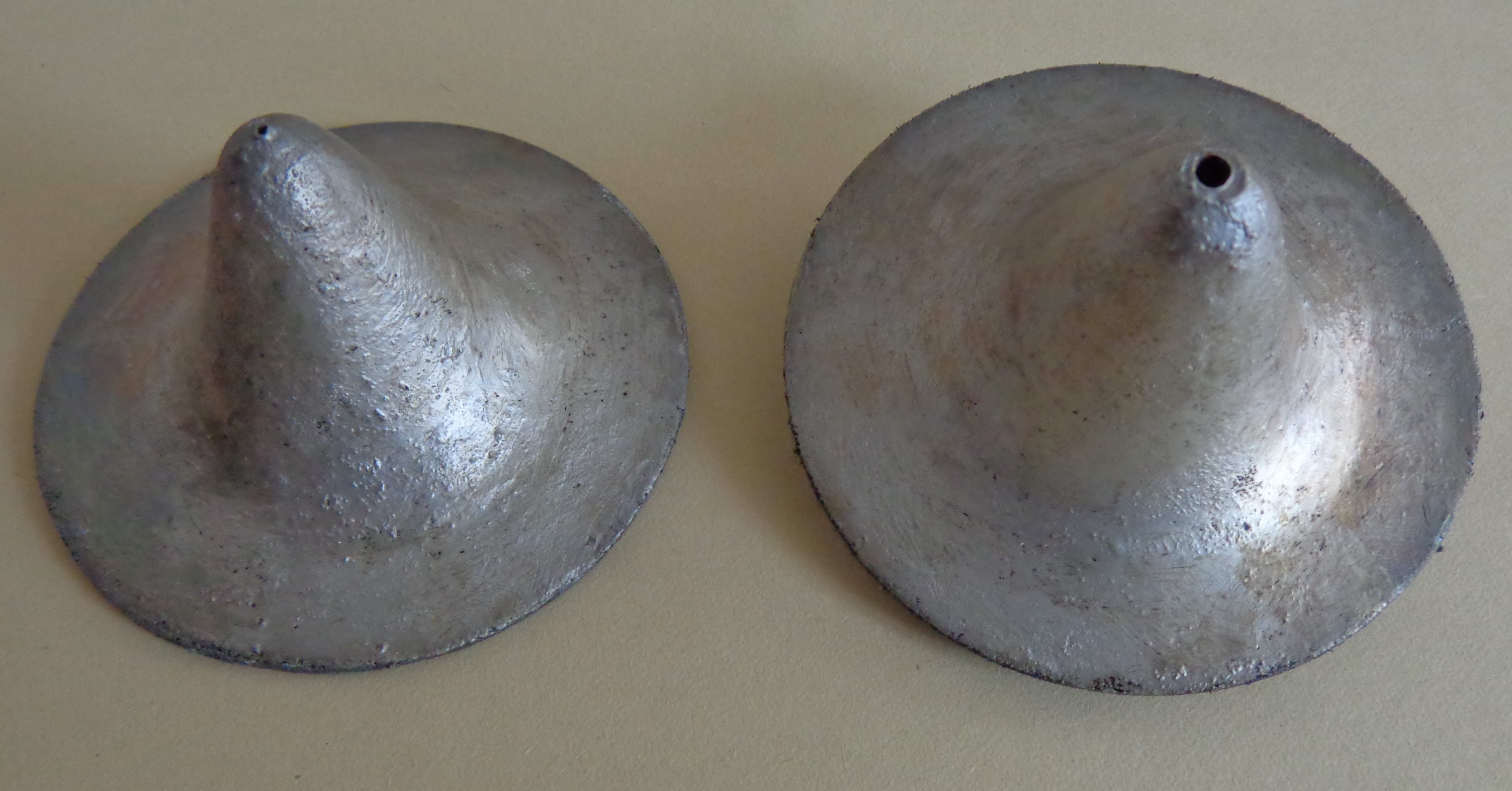}\hspace{.7cm}
\includegraphics[width=3.1cm,height=4.cm]{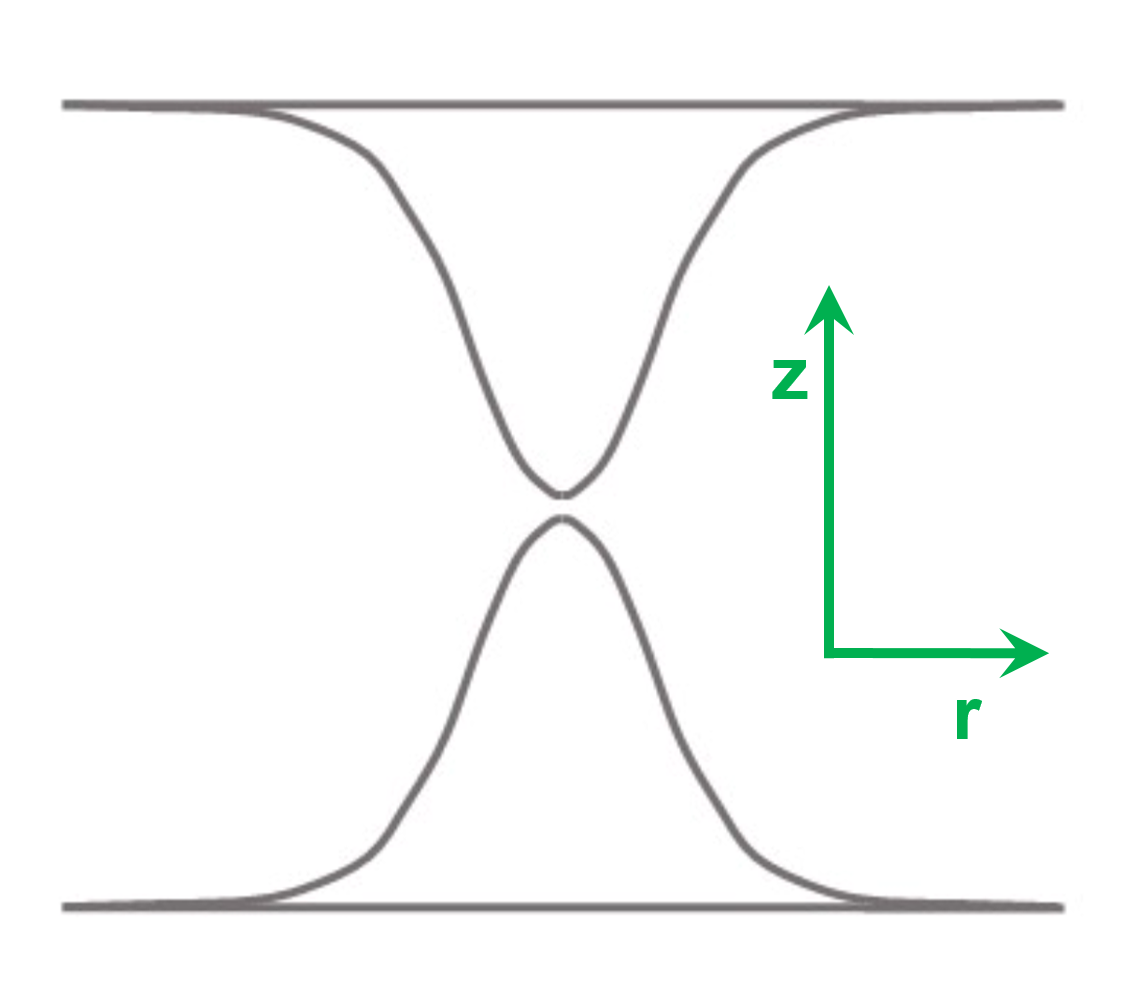}
\hspace{.2cm}
\caption{\small Type D bare biconical antenna.}\label{Fig:cn}
\end{figure}

Another set of biconical antennas was tested, using open air instead of the dielectric lens.
The copper cones were replaced by two conductors having a complicated shape, so skipping the
construction of the dielectric medium. 
3D printed antennas were fabricated using insulating materials (antenna of Type
C, made of PLA, with a base of 9 cm diameter and a height of 5.2 cm, see Fig. \ref{Fig:cv}), or
conductive materials (antenna of Type D, made of carbon black-compounded PLA, with a base of 9 cm
diameter and a
height of 4.5 cm, see Fig. \ref{Fig:cn}), both ultimately coated using a conductive silver enamel,
cured with hot air (90 degrees centigrade for 15 min). 

The shapes of these devices were obtained 
through heuristic arguments, with the help some theoretical considerations that we do not report here.
In order to be able to reproduce our experiments we provide the explicit expression
of the radius $r$ as a function of $z$, describing the curved profiles of the two devices.
For $0\leq z\leq \ell$, we have:
$$r(z)=\ell \left[ \frac65 -\left(1-\left(\frac{z}{\ell}\right)^\gamma\right)^{1/\gamma}\right]\qquad
\qquad\quad
r(z)=\ell \left[ 1 -\left(\frac{z}{\ell}\left(2-\frac{z}{\ell}\right)\right)^{1/4}\right]$$
where $\ell$ is a scaling parameter and $\gamma=1.9$.

\vspace{.5cm}

\vspace{.2cm}
\vspace{.5cm}

\begin{figure}[h!] 
\centering{
   {\includegraphics[scale=.24]{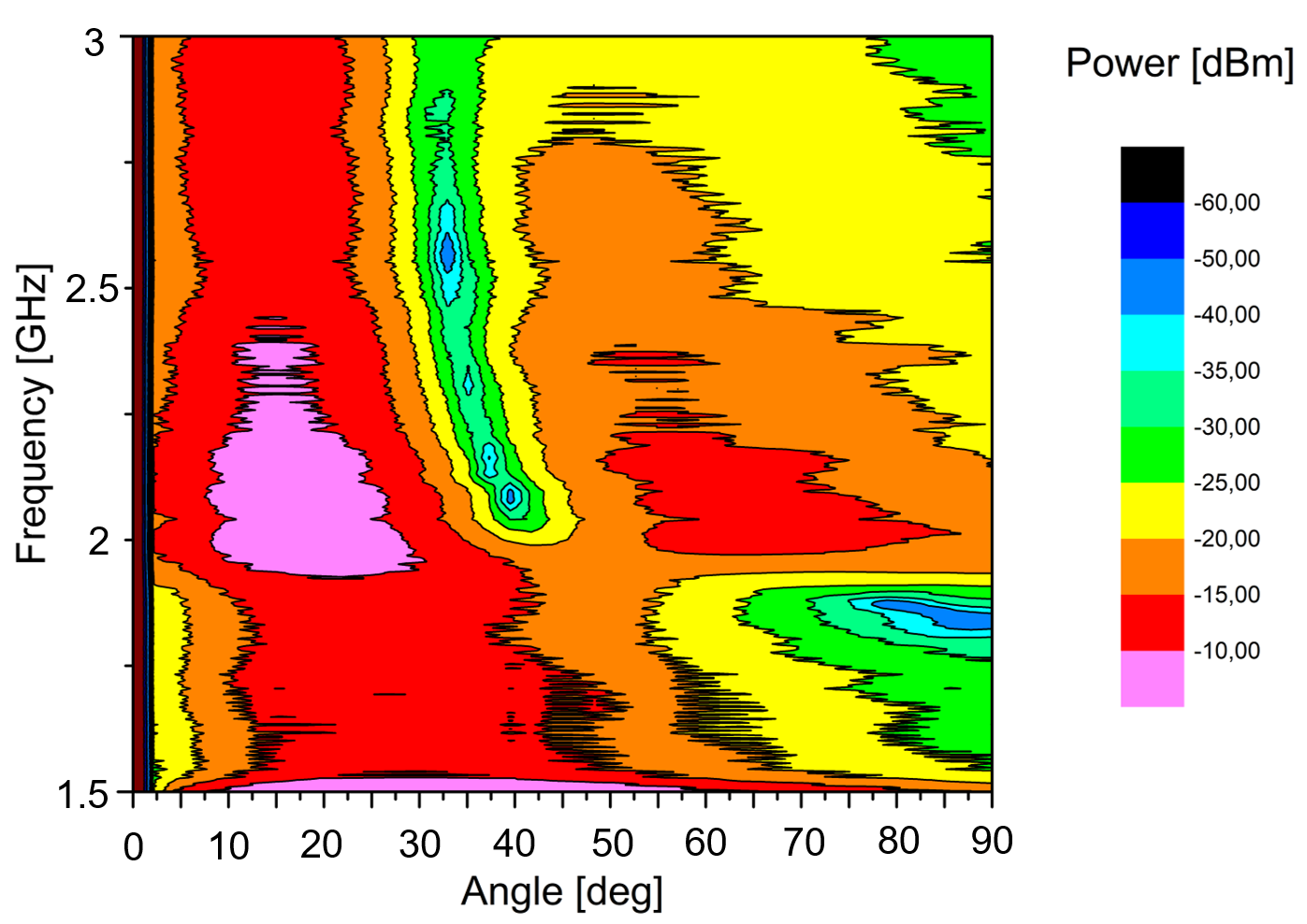}}\hspace{.2cm}
  { \includegraphics[scale=.24]{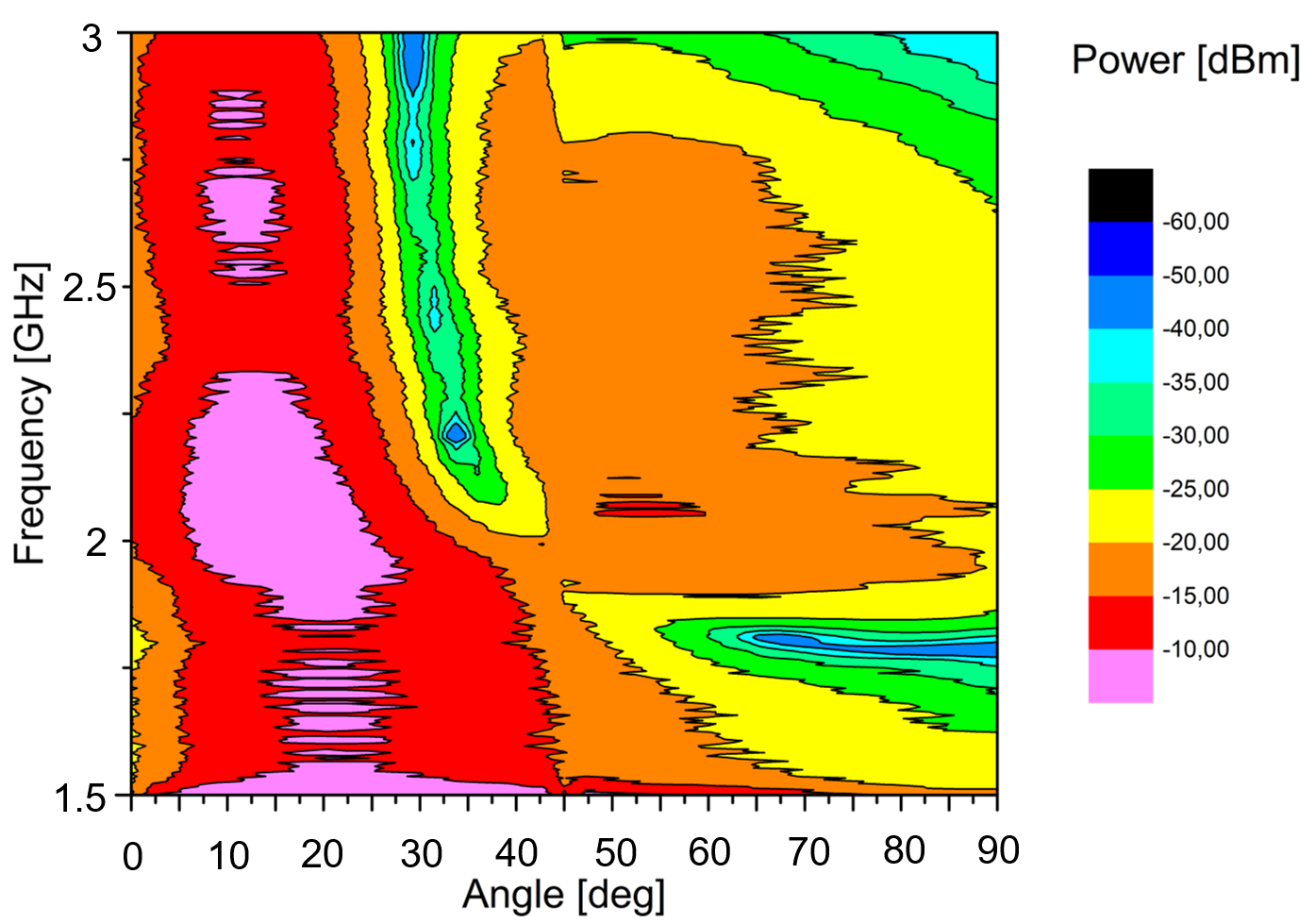}}
    }
    \vspace{.2cm}
    \caption{2D plots of bare biconical antenna emission power in dBm, recorded with a log-periodic receiver: 
    Type C (left), Type D (right).
    The frequency ranges from 1.5 GHz up to 3 GHz.
    }
    \label{fig:2Dcv}
\end{figure}

\begin{figure}[h!]
\centering{
   {\includegraphics[scale=.2]{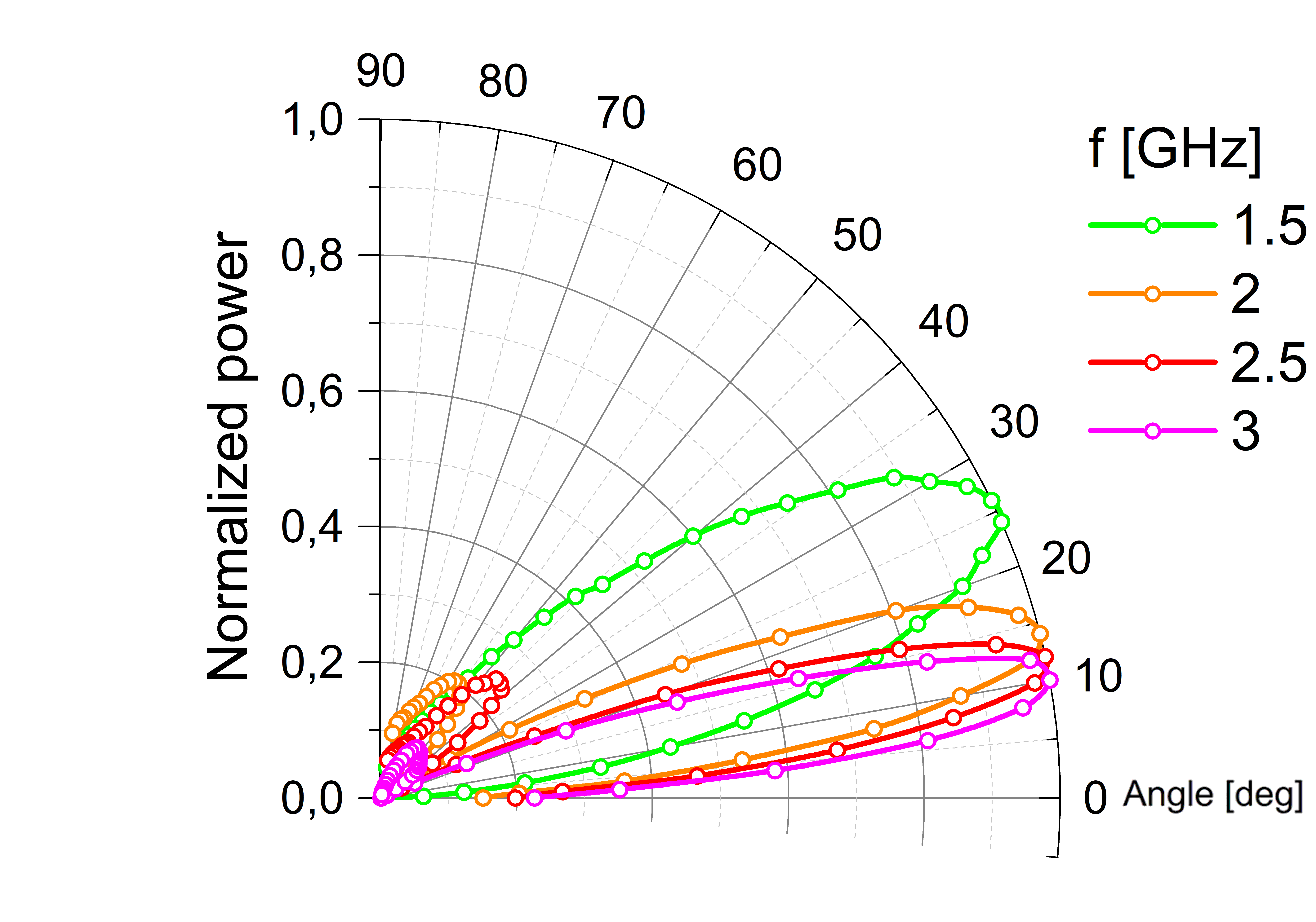}}\hspace{.0cm}
  { \includegraphics[scale=.2]{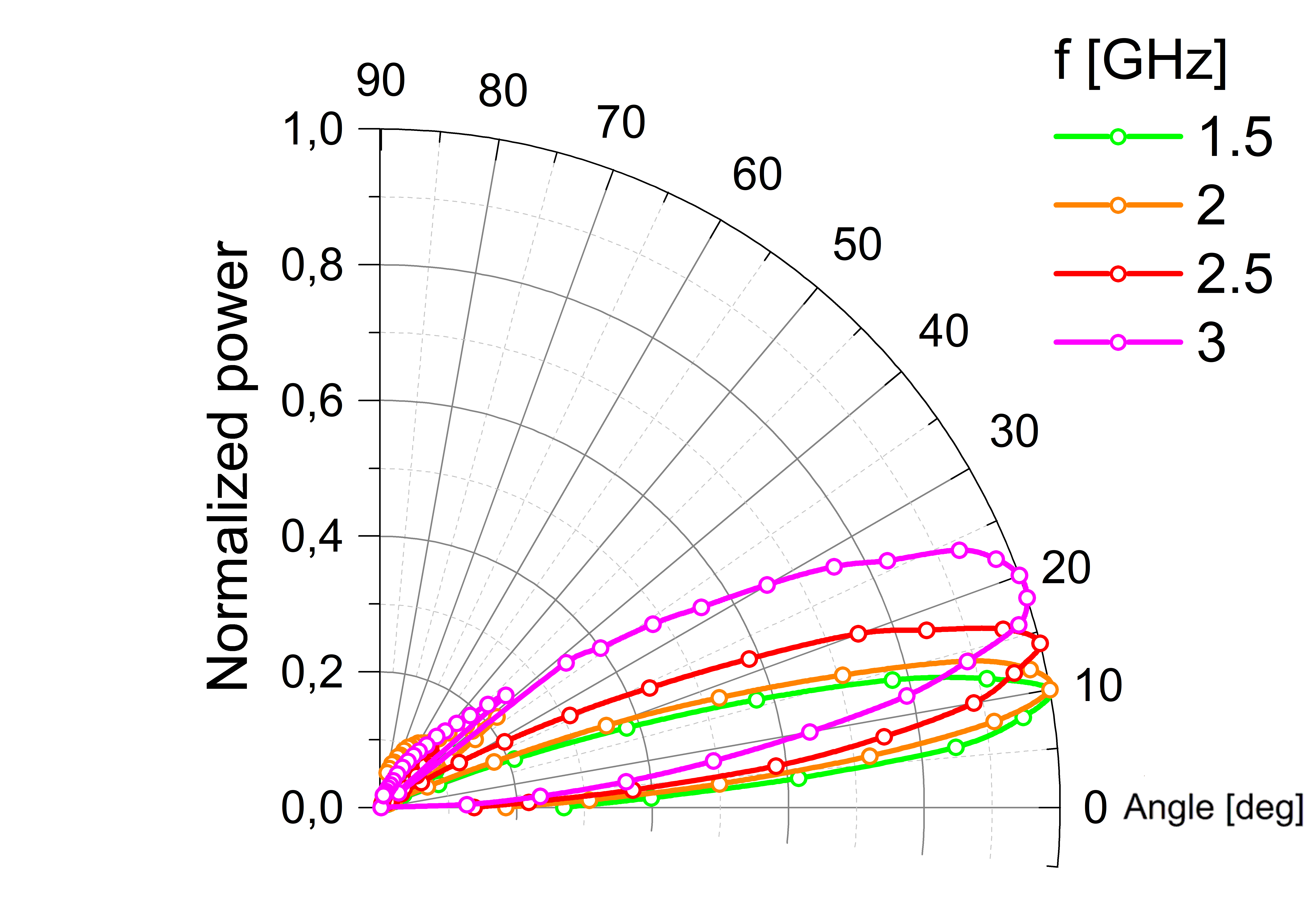}}
    }
    \caption{\small Polar diagrams of the normalized emission powers at selected frequencies of 1.5, 2, 2.5, and 3 GHz, recorded with the log-periodic receiver: Type C (left), Type D (right).
    }
    \label{fig:polarmapsOLD}
\end{figure}

Measurements have been done in the same anechoic chamber of the tests of the previous
section using the log-periodic receiver and a
 frequency range  between 1.5 GHz and 3 GHz. Figure \ref{fig:2Dcv} shows the 2-dimensional
 contour plots of the emission pattern measured by varying the angle from $0^\circ$ to $80^\circ$, whereas Figure \ref{fig:polarmapsOLD} shows the polar curves for some selected frequencies.
 Both types of transmitting antennas
 show a primary and secondary lobe emission between 2 GHz and 3 GHz, and an increasingly higher directivity
 at the higher frequencies. The antenna of Type C features a slightly better directivity and 
 works well at lower frequencies, in comparison to the device of Type D.
 The optimal directivity was found for the frequency of 3 GHz for both types of antennas, with a lobe 
 centered around $10^\circ$ featuring a FWHM of $14.2^\circ$ (Type C) and a lobe centered around $18^\circ$ featuring a FWHM of $26.3^\circ$ (Type D).
The outcome is interesting, though still a bit far from optimality.
Due to the lack of the dielectric lens correction, 
the wave grows with distance from the source, following a conical pattern.

\vspace{.3cm}
\vspace{.5cm}


\vspace{.5cm}
\vspace{.5cm}

\section{Theoretical justifications}\label{sect:theory}

The signals produced by the new types of antennas here presented are well described
in cylindrical coordinates $(r, \phi, z)$ by the expression:
\begin{equation}\label{sol1}
{\bf E}=\Big(cf(r)\sin\omega(ct-z),0,0\Big)\ \quad 
{\bf B}=\Big(0,f(r)\sin\omega(ct-z),0\Big)\ 
\end{equation}
where $c$ is the speed of light, ${\bf E}$ and ${\bf B}$ are the electric and magnetic
fields respectively, and
the function $f$ is (almost) arbitrary. In particular, we can assign $f$ in such a way it is globally
continuous and zero outside
a prescribed domain. Like in Fig. \ref{Fig:figure2} (top), this solitary wave shifts at the
speed of light along the emission axis (the $z$-axis in this case).

The wave described in (\ref{sol1}) is only achievable in the case of infinite directivity.
More likely, we have the conical type configuration  of Fig. \ref{Fig:figure2} (bottom).
This can be described in spherical coordinates $(r, \theta, \phi)$ by the following 
exact expression:
\begin{equation}\label{sol4}
{\bf E}=\Big(0,\frac{c f(\theta )}{r}\sin\omega (ct-r),0\Big)\ \quad
{\bf B}=\Big(0,0,\frac{f(\theta )}{r}\sin\omega (ct-r)\Big)\ 
\end{equation}
Here $f$ could be allowed to be only different from zero in the circular annulus determined by
$0< \theta_{\rm min}<\theta_{\rm max}<\pi/2$. When $\theta_{\rm min}$ and $\theta_{\rm max}$
are small, we get a conical wave with a very reduced aperture.

We remark that very few of the above examples are included in the solutions space of Maxwell's equation in vacuum.
Indeed, note that in general we  have ${\rm div}{\bf E}\not =0$.
For example, in (\ref{sol1}) we can only have ${\rm div}{\bf E} =0$ when $f(r)=1/r$. This last situation is compatible
with a wave traveling in a coaxial cable in conditions of perfect conductivity. 
It is however necessary to drop the equation ${\rm div}{\bf E} =0$ if we want to
simulate a similar behavior in free space. This a consequence of the vanishing boundary conditions
to be imposed to the transverse electric field along the transition zone between
the wave itself and the open space (i.e. outside the cylinder of Fig. \ref{Fig:figure2} (top) or the
cone of Fig. \ref{Fig:figure2} (bottom)). 
 
A quick check of these statements can be done in the case of a scalar wave eqwuation in Cartesian
coordinates. We look for a solution of the type $u(x,y,z,t)=f(x,y)g(ct\pm z)$, which is traveling along the $z$-axis. 
Here $g$ is arbitrary and modulates the signal in the direction of propagation.
We have:
$$
\Box u =\frac{\partial^2 u}{\partial t^2}- c^2\Delta u=\left( \frac{\partial^2 u}{\partial t^2}
-c^2 \frac{\partial^2 u}{\partial z^2}\right)-c^2\left( \frac{\partial^2 u}{\partial x^2} +
\frac{\partial^2 u}{\partial y^2}\right)$$
\begin{equation}\label{wave}
\qquad\qquad =-c^2 g\left( \frac{\partial^2 f}{\partial x^2} +
\frac{\partial^2 f}{\partial y^2}\right) =0
\end{equation}
Therefore $f$ turns out to be harmonic as a function of the variables $x$ and $y$. Since
$f$ is required to vanish outside a bounded 2-D domain, then the only possibility is
that $f\equiv 0$ (in fact, we have an elliptic problem with zero right-hand side and zero
boundary conditions).

Note also that the equation $\Box {\bf E} =0$ is obtained under the assumption that
${\rm div}{\bf E} =0$. Indeed, we have:
\begin{equation}\label{wave2}
\frac{\partial^2 {\bf E}}{\partial t^2}=c^2 \frac{\partial }{\partial t}{\rm curl}{\bf B}
=-c^2{\rm curl}\Big( {\rm curl}{\bf E}\Big)=c^2\Big[ \Delta {\bf E}-\nabla ({\rm div}{\bf E})
\Big] =c^2 \Delta {\bf E}
\end{equation}
Let us observe instead that it is always possible to arrange the distribution of the fields 
in such a way that ${\rm div}{\bf B} =0$. We just saw above that the solutions of the
wave equation are very few. Hence, a spectrum of possible solutions is brought by admitting
the possibility that ${\rm div}{\bf E} \not=0$ in suitable regions of space. As a final
remark, we may observe that the set of Maxwell's equations in vacuum amounts to eight
equations (two vector equations plus two scalar ones) with six unknowns (the components
of the electric and magnetic fields), so that an incompatibility is expected.
The possibility that ${\rm div}{\bf E} \not=0$ may sound
unphysical, since there is not presence of charges in vacuum. By the way is very useful
from the technical viewpoint, and has been a source of ideas in the development 
of the device described in this paper.

We deduce that the electromagnetic signals measured in the tests performed so far have no representation
in the usual context (see Fig. \ref{Fig:ruota}, right). One of the reasons is that the
solutions of the 3D vector wave equation tend to be dispersive in a very short range (usually
no more than one or two wavelengths), unless they
display infinite energy \cite{sezginer}. In our experiments the distance from the antenna under test
and the receiver allowed for 12 up to 60 wavelengths (depending on the signal frequency, whose
range varied from 1 GHz to 5 Ghz). Thus, the signal measured cannot belong to those modeled by 
the classical Maxwell's equations. This observation is important because the existence of the waves generated by 
our antennas is not trivially covered by standard theories. This motivates the study of suitable
extensions of the electrodynamics equations.
Many authors have devoted their efforts with the aim of detecting valuable alternatives. 
Existing literature is very rich. In  this regard we just mention a few papers: 
\cite{arbab}, \cite{benci2}, \cite{coclite}, \cite{donev}, \cite{harmuth}, \cite{hively}
\cite{lehnert},\cite{munz}.  

Since the pioneering paper \cite{bornin} (see also \cite{yang}), modifications are usually obtained by proposing an alternative Lagrangian.
Other possible justifications for the experiments described in this paper can be found in the material collected
in \cite{Fun2} and \cite{Fun3} (see the appendices there for a quick review).
There the Lagrangian is exactly the one of the usual Maxwell's equations, but the variations are subject
to a further constraint that forces the solutions to follow the rules of geometrical optics.
In this fashion one can build solutions as in (\ref{sol1}) and (\ref{sol4}), where the Poynting
vector is perfectly lined up with the direction of motion. This property holds for 
plane waves of infinite extension, though is rarely found in the Maxwellian context (see for instance the 
case of spherical waves in the near-field regime \cite{jackson}).

We just mention the version of that model that deals with the so called
{\sl  free-waves}, being a complete discussion outside the scopes of the present paper. 
This reads as follows:
\begin{equation}\label{sfem2}
\frac{\partial {\bf E}}{\partial t}~=~ c^2{\rm curl} {\bf B}~
-~{\bf V}{\rm div}{\bf E}
\end{equation}
\begin{equation}\label{sfbm2}
\frac{\partial {\bf B}}{\partial t}~=~ -{\rm curl} {\bf E}
\end{equation}
\vspace{-.2cm}
\begin{equation}\label{sfdb2}
{\rm div}{\bf B} ~=~0
\end{equation}
\begin{equation}\label{slor1}
({\bf E}+{\bf V}\times {\bf B} )~{\rm div}{\bf E}~=~0
\end{equation}
where ${\bf V}$ is a suitable velocity field associated to the flow
of energy of an electromagnetic wave.
The system admits solutions where the vectors ${\bf E}$, ${\bf B}$, ${\bf V}$ form 
orthogonal triplets. In this way, ${\bf V}$ is lined up with the  Poynting's vector given by
${\bf E}\times {\bf B}$.
The expression in (\ref{sol1}) is now solution by setting ${\bf V}=(0,0,c)$, 
and for the one in (\ref{sol4}) we have ${\bf V}=(c,0,0)$.
In this context, the Gauss theorem still holds but should be interpreted in weak form
by assuming that wells and sources compensate along the path of a wave. This fact can
be checked after a direct computation on the exact solutions.
Of course, if ${\rm div}{\bf E}=0$, the field ${\bf V}$ disappears and we return to the classical 
Maxwell's equations in vacuum. Note that the fields (\ref{sol1}) and (\ref{sol4})
solve a 1-D wave equation in the direction of propagation ${\bf V}$, though they
occupy a volume in the transverse direction. Therefore, the great advantage of this approach is to be
able to deal with a 1-D scalar wave equation in place of a 3-D vector one. According to P. Dirac,
if we substitute $f$ in (\ref{sol1}) and (\ref{sol4}) by a {\sl delta} function, we are following
the trajectory of a particle (photon) along an optical ray line. This passage has however
no explicit mathematical justification, since such a distribution would not satisfy the 
whole set of eight Maxwell's equations in vacuum. It is instead compatible with the set of equations
(\ref{sfem2}), (\ref{sfbm2}), (\ref{sfdb2}) and (\ref{slor1}).

The last observations imply that the revised set of equation is perfectly compatible with the rules of geometrical optics. 
As a matter of fact,
when ${\bf V}$ is the gradient of a potential $\Psi$ and its intensity is constantly equal to $c$,  we obtain 
$\vert {\bf V}\vert=\vert\nabla \Psi\vert =c$, which is the stationary eikonal equation
(see \cite{bornin}, section 3.1.4).
 Such a property is independent of the signal carried and describes the 
evolution of electromagnetic wave-fronts by optics. 
Moreover, the set of equations (\ref{sfem2})-(\ref{slor1}) derives from a Lagrangian,
it is invariant under Lorenz transformations, and can be written in covariant form.
This theoretical approach actually inspired the present research 
on antennas of a particular shape. The signals so generated confirm the predictions, 
also suggesting how the results can be possibly improved.

\section{Further options and conclusions}\label{sect:options}

We collected here a series of experiments aimed at showing that antennas of extremely 
high directivity can be actually built. The electromagnetic waves so produced have the 
property of preserving the topology of the magnetic field during the entire process 
that goes from the transmission of the signal inside the supplying coax cable up to 
the  emission in open space. Our experiments actually confirm the existence of such waves. 
Conical type waves are similar to that shown in  Fig.  \ref{Fig:figure2}, bottom.
We strongly believe that emission patterns exhibiting infinite directivity 
(Fig.  \ref{Fig:figure2}, top) could be also generated with the help of the due technical 
effort. Indeed, we have all the elements to ameliorate the results obtained so far. 
The reader can easily realize that achieving perfect directivity would
have a deep impact in many applications, in particular those requiring secure connections 
based on a high level of reliability (consider for instance banks, private companies, or military uses). 
In this way, RF data can be transferred between two stations through a beam that cannot 
be intercepted outside the segment
joining the two parts.

The tests we made were based on two different strands. There are in fact experiments where the 
resonator is composed by opposite cones and the system of the two is embedded in a suitable 
dielectric lens. As we checked, performances are influenced by the shape of the dielectric medium. 
There are also documented experiments where the dielectric is absent, and the resonators feature 
a peculiar shape (see Figs. \ref{Fig:cv},
\ref{Fig:cn}). Therefore, there is the possibility of combining these two approaches and play 
with ``custom'' antennae, where one can freely vary both the conductive part and the dielectric. 
The use of 3D printing techniques becomes then crucial.

\vspace{.5cm}
\begin{figure}[h!]
\centering
\includegraphics[width=5.5cm,height=4.2cm]{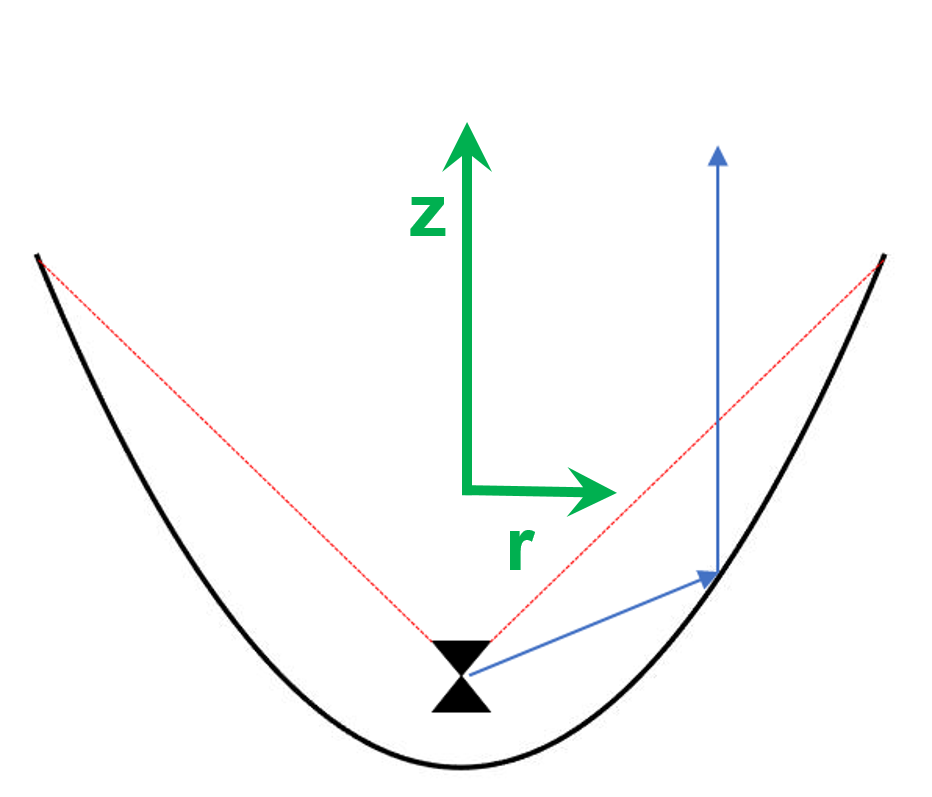}\hspace{.3cm}
\includegraphics[width=6.1cm,height=4.1cm]{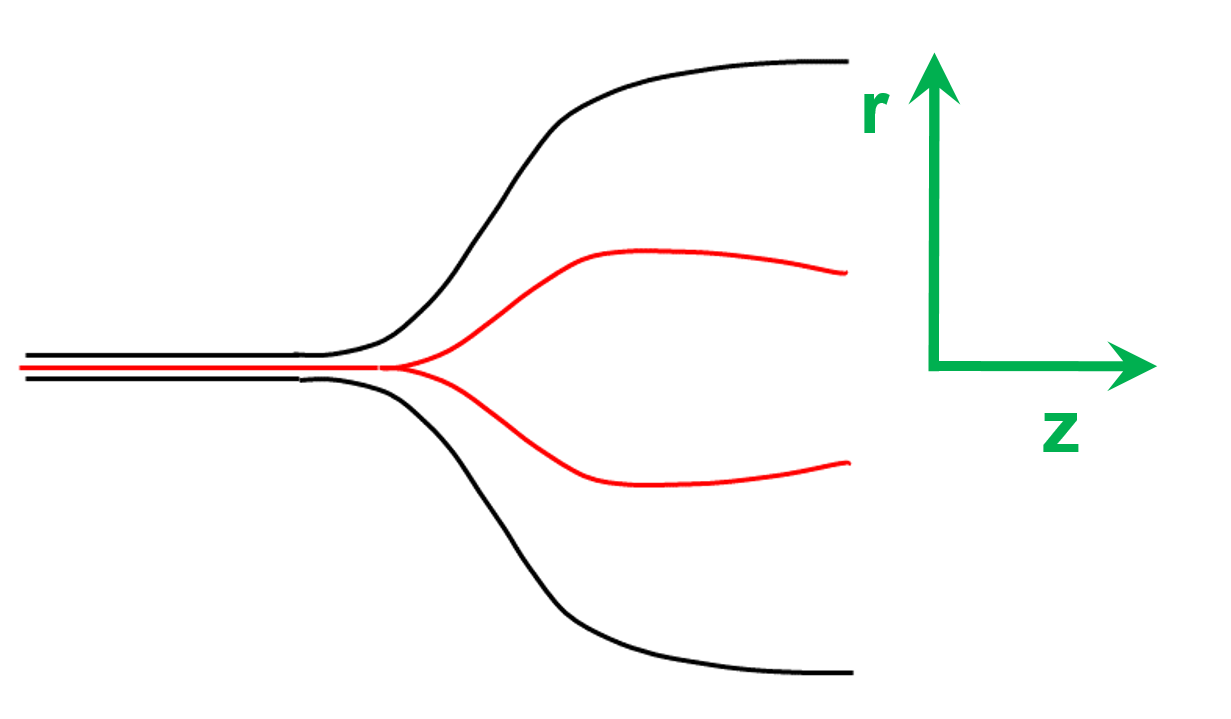}
\caption{\small Left: A directive antenna where the biconical part lays at the
focus of a large paraboloid. The displacement is rather different from the classical one, 
where the focus usually hosts a dipole whose orientation is not aimed
at preserving the topology of the magnetic lines of force. Indeed, the break of these
lines jeopardizes the directivity properties we are discussing in the present paper.
Right: Possible design of an antenna where the signal from
a coax passes through a resonating chamber before being definitively emitted.
The 3D body of both antennas is obtained by rotating the pictures
about the emission axis (z).}
\label{Fig:figure8}
\end{figure}
\vspace{.5cm}

There is in addition a third pathway, which consists of exploring other reflector types.
Here we used a conical reflector due to its simplicity of construction. Nevertheless, a parabolic type
reflector could work as well (see Fig. \ref{Fig:figure8}, left). In this fashion,
the biconic resonator must be put at the focus. The disadvantage is that the device
becomes rather bulky. We could also take into account situations where the reflector
is removed (see Fig. \ref{Fig:figure8}, right). Here the geometry becomes extremely
decisive. The aim is always to preserve the circular topology of the lines of force of 
the magnetic field. The magnetic induction streamlines of the signal coming from the coax 
are closed transverse circles. These are somehow ``dilated'' when passing through the antenna device.
If some conditions of resonance are met (strictly depending on shape and frequency), 
the wave could be expelled in open air. We have no idea however if such last experiment is feasible,
though some preliminary tests have been already tried with little success.

We mention some negative effects that should be better investigated. One of them is the design
of the area involved in the signal input. The gap between the cones must be as tiny as possible.
In the geometries we tested here, the cones were too small to neglect this factor.
This choice also forced us to work at higher frequencies, with the consequence of being more
exposed to the various imperfections of the prototypes. In addition, the solid printing
of the dielectric in PLA, in the case of Type B, took many hours. Therefore, devices of
larger size come at a nontrivial cost. When the input frequency is not high enough,
the antenna does not resonate in the proper manner. The signal can escape from the edges
at the base of the cones, generating unwanted effects. This could be an explanation of the
appearance of secondary lobes at lower frequency (the spurious wave is directly radiated,
without passing through the dielectric and the reflector). Such a fact is certainly true for
the devices of Type C and D, where at certain angles the gap between the conductors
is directly visible at naked eye, which implies that there are emitted rays that do not
encounter the reflector.
Notwithstanding the roughness of our first tests, we are proud to claim that the results 
are not bad at all.

Of course, we may elaborate  more complicated situations
were all the degrees of freedom (i.e., the shapes of the resonator, the dielectric
and the reflector) are taken into consideration. These are the reasons why
we are convinced that, one day, antenna devices displaying infinite directivity will
become a reality. Point-to-point communications without dispersion can be then 
set up by using two antennas of  the same kind both for transmission
and reception.

\end{document}